\documentclass[aps,prl,reprint,superscriptaddress,twocolumn,showpacs,amsmath,amssymb, amsfonts,10pt]{revtex4-2}
\usepackage[T1]{fontenc}
\usepackage{graphicx}
\usepackage{enumitem}
\usepackage{amsmath}

\usepackage{amsfonts}

\usepackage{amsmath}
\usepackage{amsfonts}
\usepackage{amssymb}
\usepackage{mwe}
\usepackage{graphicx}
\usepackage{color}
\usepackage{bm}
\usepackage{soul}
\usepackage[normalem]{ulem}

\usepackage[colorlinks,bookmarks=false,urlcolor=blue, citecolor=blue, linkcolor = blue]{hyperref}

\begin{document}
\title{Superexchange liquefaction of strongly correlated lattice dipolar bosons}
\author{Ivan Morera}
\affiliation{Departament de F{\'i}sica Qu{\`a}ntica i Astrof{\'i}sica, Facultat de F{\'i}sica, Universitat de Barcelona, E-08028 Barcelona, Spain}
\affiliation{Institut de Ci{\`e}ncies del Cosmos, Universitat de Barcelona, ICCUB, Mart{\'i} i Franqu{\`e}s 1, E-08028 Barcelona, Spain}
\author{Rafa\l{} O\l{}dziejewski}
\email[Correspondence to: ]{rafal.oldziejewski@mpq.mpg.de}
\affiliation{Max Planck Institute of Quantum Optics, 85748 Garching, Germany}
\affiliation{Munich Center for Quantum Science and Technology, Schellingstrasse 4, 80799 Munich, Germany}
\author{Grigori E. Astrakharchik}
\affiliation{Departament de F\'isica, Universitat Polit\`ecnica de Catalunya, 
Campus Nord B4-B5, E-08034 Barcelona, Spain}
\affiliation{Departament de F{\'i}sica Qu{\`a}ntica i Astrof{\'i}sica, Facultat de F{\'i}sica, Universitat de Barcelona, E-08028 Barcelona, Spain}
\affiliation{Institut de Ci{\`e}ncies del Cosmos, Universitat de Barcelona, ICCUB, Mart{\'i} i Franqu{\`e}s 1, E-08028 Barcelona, Spain}
\author{Bruno Juli{\'a}-D{\'i}az}
\affiliation{Departament de F{\'i}sica Qu{\`a}ntica i Astrof{\'i}sica, Facultat de F{\'i}sica, Universitat de Barcelona, E-08028 Barcelona, Spain}
\affiliation{Institut de Ci{\`e}ncies del Cosmos, Universitat de Barcelona, ICCUB, Mart{\'i} i Franqu{\`e}s 1, E-08028 Barcelona, Spain}

\begin{abstract}
We propose a mechanism for liquid formation in strongly correlated lattice systems. The mechanism is based on an interplay between long-range attraction and superexchange processes. As an example, we study dipolar bosons in one-dimensional optical lattices. We present a perturbative theory and validate it in comparison with full density-matrix renormalization group simulations for the energetic and structural properties of different phases of the system, i.e., self-bound Mott insulator, liquid, and gas. We analyze the non-equilibrium properties and calculate the dynamic structure factor. Its structure differs in compressible and insulating phases. In particular, the low-energy excitations in compressible phases are linear phonons. We extract the speed of sound and analyze its dependence on dipolar interaction and density. We show that it exhibits a non-trivial behavior owing to the breaking of Galilean invariance. We argue that an experimental detection of this previously unknown quantum liquid could provide a fingerprint of the superexchange process and open intriguing possibilities for investigating non-Galilean invariant liquids.
\end{abstract}
\maketitle

\textbf{Introduction.} 
Ultracold atoms in optical lattices might serve as a quantum simulator of the Hubbard model that plays a vital role in our understanding of strongly correlated solid-state materials~\cite{lewenstein2012ultracold}. Particularly, the Hubbard model at strong coupling displays superexchange processes which can be employed to simulate different quantum magnetic systems~\cite{Duan2003Spin,Bloch2008Spin,Bloch2012Spin,Bloch2013Spin,Ketterle2020Spin}. Moreover, the recent experimental progress with atoms possessing strong magnetic moments, like Dy, has led to the first realization of an extended Bose-Hubbard model (eBH) in 3D~\cite{baier2016extended} and 1D~\cite{natale2022strongly}. The eBH phase diagram remains vastly uncharted. The addition of a long-range potential to the interplay between on-site repulsion and periodic confinement holds promise for exciting new physics to emerge, e.g. quasi-localization~\cite{Barbiero2015,Li2020} and unexpected topological phases~\cite{ kraus2021quantum,Barbiero2022}. 

The classical van der Waals theory of fluids states that self-bound liquids exist due to the attractive finite-range part of the interparticle interaction stabilized by the repulsive short-range core~\cite{van2004continuity}. Recently, a novel paradigmatic quantum liquid has been observed in ultracold systems that is simultaneously ultradilute and coherent~\cite{Schmitt2016,Barbut2016,Chomaz2016,Barbut2018,Boettcher2019,Tarruell2018,Tarruell2018a,Fattori2018}. Contrary to van der Waals mechanism, the weakly-interacting quantum gases undergo liquefaction due to the zero-point energy fluctuations~\cite{Petrov2015} --- the so-called Lee-Huang-Yang (LHY) term~\cite{LHY1957,Lima2011,Lima2012}. Notably, quantum liquids are even more robust in lower dimensions owing to an enhanced role of quantum fluctuations for both two-component and dipolar gases~\cite{Petrov2016,Santos2017,Malomed2018,Zin2018,Ilg2018,Rakshit2019,Oldziejewski2020,de2022,Guijarro2022}, while in classical systems the van der Waals mechanism cannot prevent a collapse of the classical system with long-range interactions in lower dimensions. Interestingly, microscopic foundations of quantum liquid formation for weakly-interacting Bose-Bose mixtures can be studied in one-dimensional optical lattices~\cite{Morera2020,Morera2021}.
\begin{figure}[t!]
        \centering
        \includegraphics[width=0.9\columnwidth]{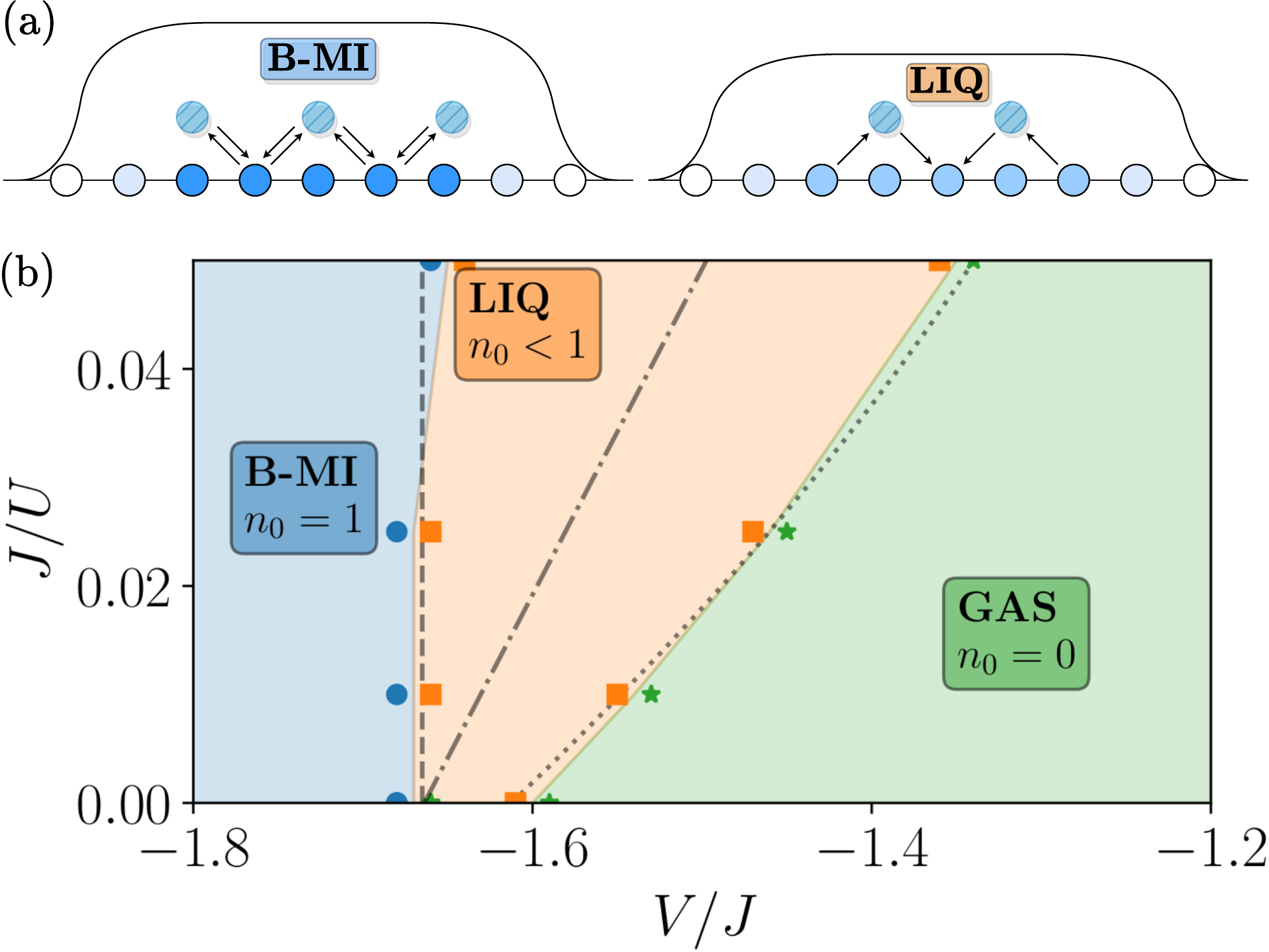}
        \caption{(a) Schematic representation of a density profile in self-bound Mott insulator (B-MI) having a unit filling and liquid drop (LIQ) in which particles are bound but holes are present. Arrows indicate the different superexchange processes through doublon formation. (b) Phase diagram for dipolar bosons in a one-dimensional optical lattice. Different phases can be characterized by their respective equilibrium densities $n_0$. We encounter a gaseous phase (GAS), a liquid one, and a self-bound Mott-insulator. Dashed and dotted-dashed lines denote respectively LIQ-to-B-MI and GAS-to-LIQ transitions obtained in perturbation theory. The dotted line indicates the threshold for a bound state (dimer) in the two-body problem.}
        \label{Fig:PhasDiagr}
\end{figure}
In this Letter, we study an unconventional microscopic mechanism of liquid formation for a system confined to a one-dimensional optical lattice with strong on-site repulsion --- in the vicinity of the Tonks-Girardeau limit (TG) --- and long-range attraction. Notably, we show that the appearance of superexchange processes liquefies the gas state close to an insulator phase transition rendering a quantum droplet for moderate and large system sizes. 
Furthermore, we study the dynamic structure factor and show that it differs substantially in compressible and incompressible phases. 
In gas and liquid phases, the excitation spectrum is exhausted by two-particle excitation with linear low-momentum dispersion, justifying the applicability of the Luttinger Liquid theory. Instead, in the insulating phase, a gap opens, and doublon excitations become dominant.
Finally, we study the speed of sound of the system. We observe changing behavior as a result of the interplay between interactions and the breaking of the Galilean invariance in lattice systems. Our theoretical predictions should be applicable to a vast range of physical systems realizing eBH models like dipolar bosons~\cite{Kao296,natale2022strongly}, Rydberg atoms~\cite{labuhn2016tunable,browaeys2020manybody} or even excitonic platforms~\cite{lagoin2022checkerboard}.

\textbf{Setup and model.}
We consider a one-dimensional (1D) system of ultracold dipolar bosons loaded to a deep optical lattice comprising $N_s$ sites. The corresponding Hamiltonian is given by an extended Bose-Hubbard (eBH) model,
\begin{align}
\hat{H}\!=\!\!\sum_{i=1}^{N_s}\!
\left[\frac{U}{2}\hat{n}_i\left(\hat{n}_i\!-\!1 \right)-J\left(\hat{b}_i^{\dagger} \!\hat{b}_{i+1}\!+\!\text{h.c.} \right)\right]  
+ V\!\sum_{i<j}^{N_s}\!\!\frac{\hat{n}_i \hat{n}_j}{|i\!-\!j|^3},
\label{Eq:BH}
\end{align}
with hopping $J$, on-site interaction $U$ and dipolar strength $V$. Importantly, it is possible to tune $U/J$ and $V/J$ separately in experiments with 1D optical lattices~\cite{Kao296}, e.g. by using Feshbach resonances and adjusting the polarization angle of the magnetic moment in the system. Thus, all the phases considered should be accessible for future experiments with Dy or Er atoms similar to existing state-of-art settings~\cite{natale2022strongly,Kao296}, see the Supplemental Material~\cite{Suppl}.
A typical transverse length of a trapping potential in current experiments with ultracold gases $\sigma_{\perp}\approx 50$~nm~\cite{Kao296} is much smaller than a typical lattice spacing $a\approx 500$~nm. We thus use pure dipolar interaction in 1D instead of the effective potential for quasi-1D geometries~\cite{deuretzbacher2010ground}, as the latter includes corrections only at short distances of the order of $\sigma_{\perp}$. Hereafter, we fix the length scale $a=1$ to the lattice spacing. 
In the following, we study the ground state of the system described by Hamiltonian~\eqref{Eq:BH} for large values of the on-site repulsion ($U/J \gg 1$) by developing perturbation theory and performing numerical simulations.

\begin{figure}[t!]
\centering
\includegraphics[width=1\columnwidth]{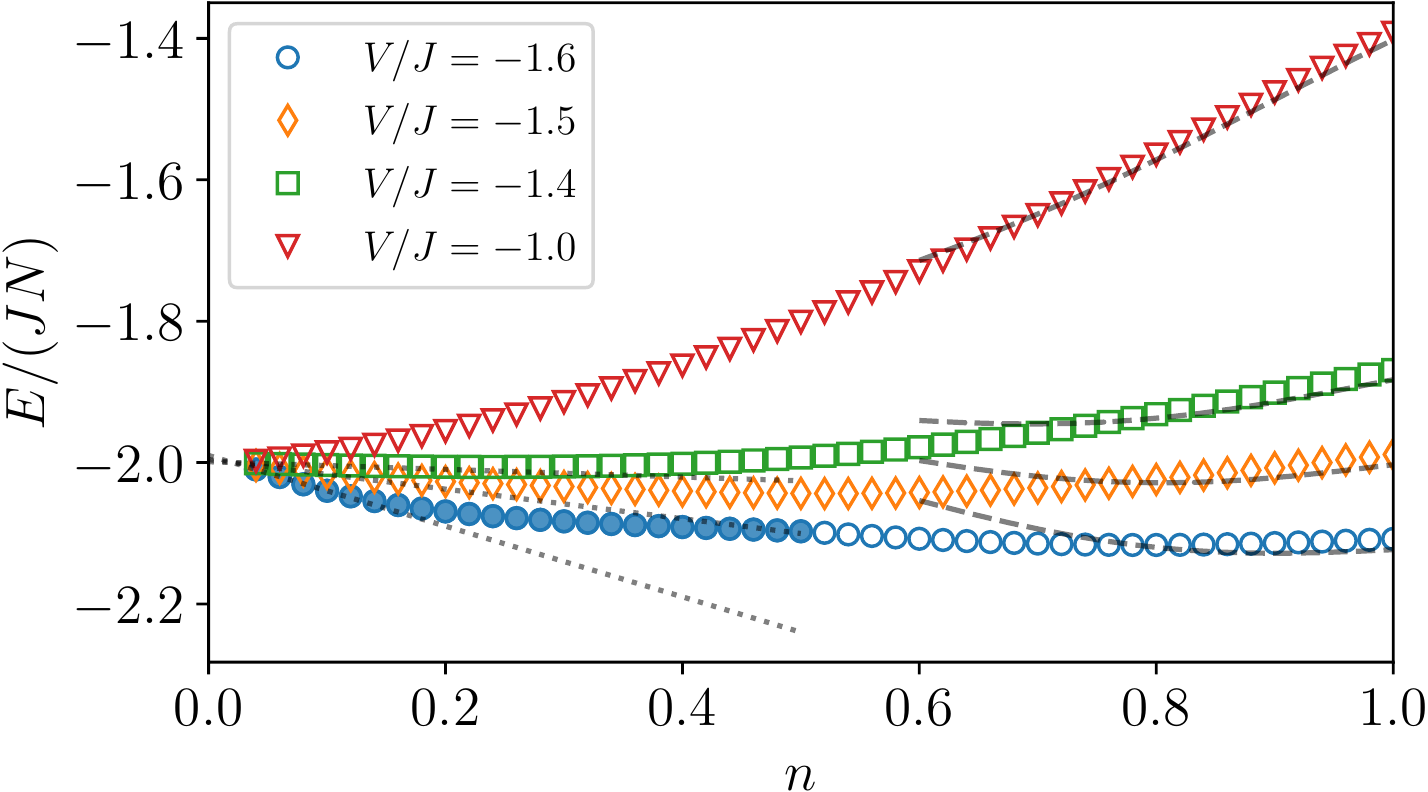}
\caption{Equation of state of dipolar bosons in a one-dimensional optical lattice for different dipolar strength $V/J$ and fixed on-site repulsion $U/J=20$. Filled symbols denote inhomogeneous solutions. Dashed lines show the Tonks-Girardeau perturbative result~\cite{Suppl}. The dotted line shows the behavior expected for a lattice soliton solution~\cite{SCOTT1994Sol}. Liquid forms when the equation of state shows a minimum at some finite value of density $n_0$ with the energy per particle $E/(N_sn_0)<-2J$, see $V/J=-1.4,-1.5,-1.6$ curves. A gas occurs when the minimum is located at zero density with the energy per particle $E/N>-2J$, see the $V/J=-1.0$ curve.
}
\label{Fig:EoS}
\end{figure}
\textbf{Insulator instability.}
We start by investigating an impenetrable lattice TG gas ($U/J\rightarrow \infty$) perturbed by an attractive dipolar interaction ($|V|\sim J$). 
We develop an analytical theory by assuming a lattice TG state $|\psi_{\textrm{TG}}\rangle$ at a density $n=N/N_s$ and calculating the effective equation of state (EoS) perturbatively as $E=\langle \psi_{\textrm{TG}}| \hat{H} |\psi_{\textrm{TG}}\rangle$. The resulting energy per particle $e\equiv E/N$ reads $e = e_J + e_V$, where 
\begin{eqnarray}
    e_J &=& -2J \sin(n\pi)/(n\pi), \\
    e_V &=& V\!\Big(\!\zeta(3)n\!-\!\frac{ \zeta(5)}{2n\pi^2}\!+\!\frac{1}{4n\pi^2}
    \left[\textrm{Li}_{5}\!\left(e^{2 i \pi n}\right)\!+\!\textrm{Li}_{5}\!\left(e^{-2 i \pi n}\right)\!\right]\!\!\Big), \nonumber 
\end{eqnarray}
indicate the kinetic energy of the fermionized bosons and the dipolar interaction energy accordingly. We also introduce the polylogarithm function $\textrm{Li}_{\beta}(n)$ of order $\beta$.

\begin{figure}[t!]
\centering
\includegraphics[width=1\columnwidth]{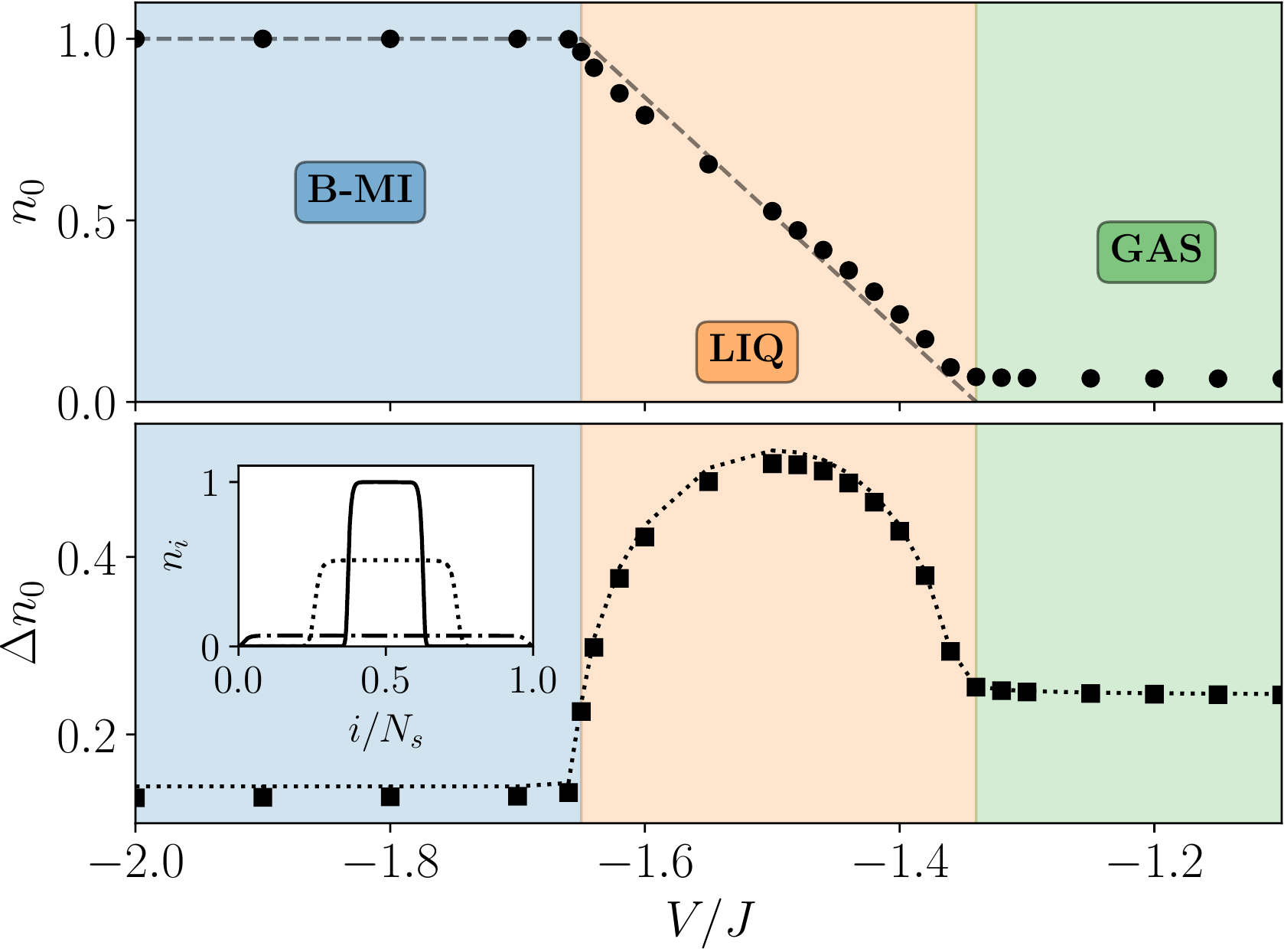}
\caption{Density and its variance as a function of $V/J$ for a fixed $U/J=20$.
Top: Saturated density $n_0$ (circles). Dashed line: unit filling, $n_0=1$ (B-MI); zero filling, $n_0=0$ (GAS); linear interpolation (LIQ). We have performed o.b.c numerical simulations up to a density $n=50/800$. Bottom: Particle number variance $\Delta n_0^2 = \langle \hat{n}_0^2\rangle - \langle \hat{n}_0\rangle^2$ (squares). The dotted line denotes the particle number variance obtained in a TG-like state, see the main text. Inset: Characteristic density profiles obtained for $V/J=-1.8$ (B-MI), $V/J=-1.5$ (LIQ) and $V/J=-1.3$ (GAS).}
\label{Fig:Eq}
\end{figure}

We classify the different phases of the system based on the value of the equilibrium density $n_0$, defined as the density at which the energy per particle is minimal. The gaseous (GAS) phase is characterized by a vanishing equilibrium density $n_0=0$ and it appears for $V< V_{\textrm{B-MI}}$~\footnote{Note that the vanishing equilibrium density ($n_0 = 0$) in the gas phase can be reached only in the $N_s \rightarrow \infty$ limit. Instead, in a finite size system, the gas always stays at some finite pressure. However, zero pressure can be found in finite systems for self-bound states (LIQ and B-MI) with an equilibrium density $n_0$.}. At the critical value $V=V_{\textrm{B-MI}}$, the minimum of the EoS jumps to unit density $n_0=1$ signaling a first-order transition to an insulator state. Furthermore, the insulator state has a lower energy per particle than the free particle energy $-2J$. It is thus a self-bound state, to which we refer as a self-bound Mott insulator (B-MI). The threshold of the B-MI state is defined by the condition that the energy per particle at unit filling equals the free energy per particle $e(n=1)=-2J$, giving the critical value of the long-range interaction $V_{\textrm{B-MI}}/J = -2/\zeta(3)$.  B-MI states feature complete incompressibility. Thus in finite systems they become completely localized exhibiting compact density profiles with a saturated density corresponding to strictly one particle per site $n=1$ (see Fig.~\ref{Fig:PhasDiagr}). Although the density profile for these states bears similarity to quantum droplets, quantum correlations are completely suppressed in the B-MI case (see also Fig.~\ref{Fig:Eq}). Therefore, one should not consider them as genuine liquids. In fact, they can be related to phase-separated states in spin systems~\cite{Petrosyan2007}. 

Importantly, lattice physics  significantly differs from what one expects to find in a continuous analog of the described system. In the 1D continuum, the addition of attractive dipolar interaction to TG gas leads to the collapse of the system as the quantum fluctuations cannot compensate for the diverging dipolar attraction. The feature of a lattice is that it provides a natural regularization of the problem since, for hard-core particles, the maximum density allowed is fixed by the lattice spacing ($n \leq 1$) \footnote{Note that this holds in the regime of applicability of the single-band eBH. For very strong attractive dipolar interactions, one may need to include higher bands into consideration.}. Notice, however, that genuine self-bound droplets in the TG limit for quasi-1D dipolar bosons, where the existence of transverse structure regularizes short-range divergence, were predicted recently~\cite{Oldziejewski2020}. 

\textbf{Superexchange processes and liquefaction.} 
The mechanism lying behind B-MI formation is based on a near cancellation between the effective 
repulsion coming from the kinetic energy of lattice hard-core particles and the long-range dipolar attraction. This scenario resembles a droplet formation for weakly interacting bosons with canceling mean-field contributions and stabilization due to beyond mean-field LHY terms. In the following, we study the effects of relaxing the hard-core condition in our system. 

We consider penetrable bosons with large but finite on-site interaction $U/J\gg 1$. We thus move away from the TG limit that opens the possibility of next-to-nearest neighbor hopping through a virtual intermediate process of two bosons occupying the same site --- the so-called superexchange process. Explicitly, this extra contribution $e_U$ to the total energy $e$ can be derived using the effective fermionic Hamiltonian within the second-order degenerate perturbation theory~\cite{PhysRevA.67.053606,Suppl}, and it reads:
\begin{equation}\label{Eq:EoSQTG}
    e_U = -\frac{4J^2}{U} n \left(1- \frac{\sin(2\pi n)}{2\pi n}  \right).
\end{equation}
Note that the superexchange correction effectively introduces additional attraction of order $J^2/U \ll 1$ in the system for all densities.

Remarkably, by inspecting the modified EoS of the ground state, we encounter a gas-to-liquid transition close to the B-MI boundary owing entirely to the emerging superexchange process, which diminishes on-site repulsion. 
The found liquid phase is characterized by a negative energy per particle smaller than for the free case $E/N<-2J$ and a finite equilibrium density $0<n_0<1$ (see Fig.~\ref{Fig:PhasDiagr}). By increasing the attractive long-range coupling, the density saturates with one particle per site, and consequently, the system enters the B-MI state.

Notably, the discovered mechanism of liquefaction in the strongly-correlated regime bears certain similarities with the one vastly studied in the opposite limit of weak interactions. As therein, quantum liquids appear due to the subleading terms when dominating contributions of opposite signs cancel each other. Nonetheless, contrary to the LHY term as a small beyond mean-field effect, the superexchange correlations occur for strongly interacting systems well beyond the mean-field applicability.

\textbf{Equilibrium properties.}
To benchmark the perturbative EoS, we calculate the ground-state energy of Hamiltonian~\eqref{Eq:BH} using the density matrix renormalization group (DMRG) algorithm, see the Supplemental Material for details. We compute the ground state for different strengths of the long-range interaction $V/J$ keeping fixed a large but finite on-site interaction $U/J$, see Fig.~\ref{Fig:EoS}. In the case of larger densities ($n \gtrsim 0.8$), an excellent agreement is found for any value of the long-range interaction, especially when approaching the unit filling limit $n=1$ owing to suppressed quantum correlations. For smaller densities ($n\lesssim 0.5$), the effective EoS deviates significantly, as it does not include the presence of few-body bound states. In DMRG simulations, we always observe that the appearance of the liquid phase coincides with the formation of a two-body bound state in the system, see Fig.~\ref{Fig:PhasDiagr}. Details on solving the two-body problem are given in~\cite{Suppl}.
In the few-body limit, the equation of state is well approximated by a lattice soliton solution~\cite{SCOTT1994Sol} $\frac{E}{N}+2J\sim \frac{\left(N-1\right)}{2} \epsilon_b$ with $\epsilon_b$ being the two-particle bound-state energy, see Fig.~\ref{Fig:EoS}. The appearance of the lattice soliton shows that the long-range effects of the dipolar interaction at small densities can be absorbed into the binding energy $\epsilon_b$.

The EoS allows one to differentiate three distinct phases in the system, see Fig.~\ref{Fig:EoS}. The gaseous phase features a minimum in the EoS at zero density $n_0=0$ and a positive energy per particle compared to the free case $e>-2J$. On the other hand, the liquid and B-MI phases are characterized by negative binding energies $e<-2J$ in the minimum. The equilibrium density of a liquid ranges within $0<n_0<1$, and saturates to $n_0=1$ in the B-MI phase, making its EoS singular. Moreover, the liquid state is compressible while the B-MI is not. 

One of the qualitative differences between the predictions of perturbative theory and DMRG results is the presence of a narrow liquid phase in the TG limit $J/U=0$. This liquid state is sandwiched between the gas and B-MI phases for $V_{\textrm{MI}}<V<-1.61J$, see Fig.~\ref{Fig:PhasDiagr}. The weakest attraction, for which the liquid forms, is defined by the existence of a two-body bound state. Remarkably, for the dipolar interaction, the formation of a two-body bound state does not coincide with the formation of a self-bound MI state in the many-body problem for $J/U=0$, in contrast to faster decaying potentials. We thus relate the presence of the liquid phase to the long-range nature of the dipolar interaction. 

The equilibrium properties differ dramatically among the various phases, as shown in Fig.~\ref{Fig:Eq}. The equilibrium density ranges from $0<n_0<1$ in the liquid phase and saturates to $n_0=1$ in the B-MI phase. DMRG simulations result in density profiles that do not fill the entire lattice for these phases, contrary to the gaseous state, see inset of Fig.~\ref{Fig:Eq}. Moreover, the value of the saturated density agrees excellently with the equilibrium density found from the EoS. Its value increases almost linearly for the growing attractive long-range interaction in the liquid phase. Upon reaching the B-MI phase, the equilibrium density saturates to $n_0=1$ and becomes independent of the dipolar coupling.

Another important observable pertains to the particle variance $\Delta n_0=\sqrt{\langle \langle \hat{n_0}^2\rangle-n_0^2\rangle}$, as it quantifies the fluctuations of density and probes the structure of pair correlations. Particularly, it vanishes in the gas phase (due to vanishing equilibrium density) and in the B-MI phase becomes of order $J/\sqrt{U}$.
We find that the maximal value of $\Delta n$ is reached in the liquid phase close to half-filling $n_0\sim 1/2$. Fascinatingly, the lattice TG model captures the particle variance model well and provides a precise analytic description, see Fig.~\ref{Fig:Eq}.

\textbf{Non-equilibrium properties.}
\begin{figure}[t!]
\centering
\includegraphics[width=1\columnwidth]{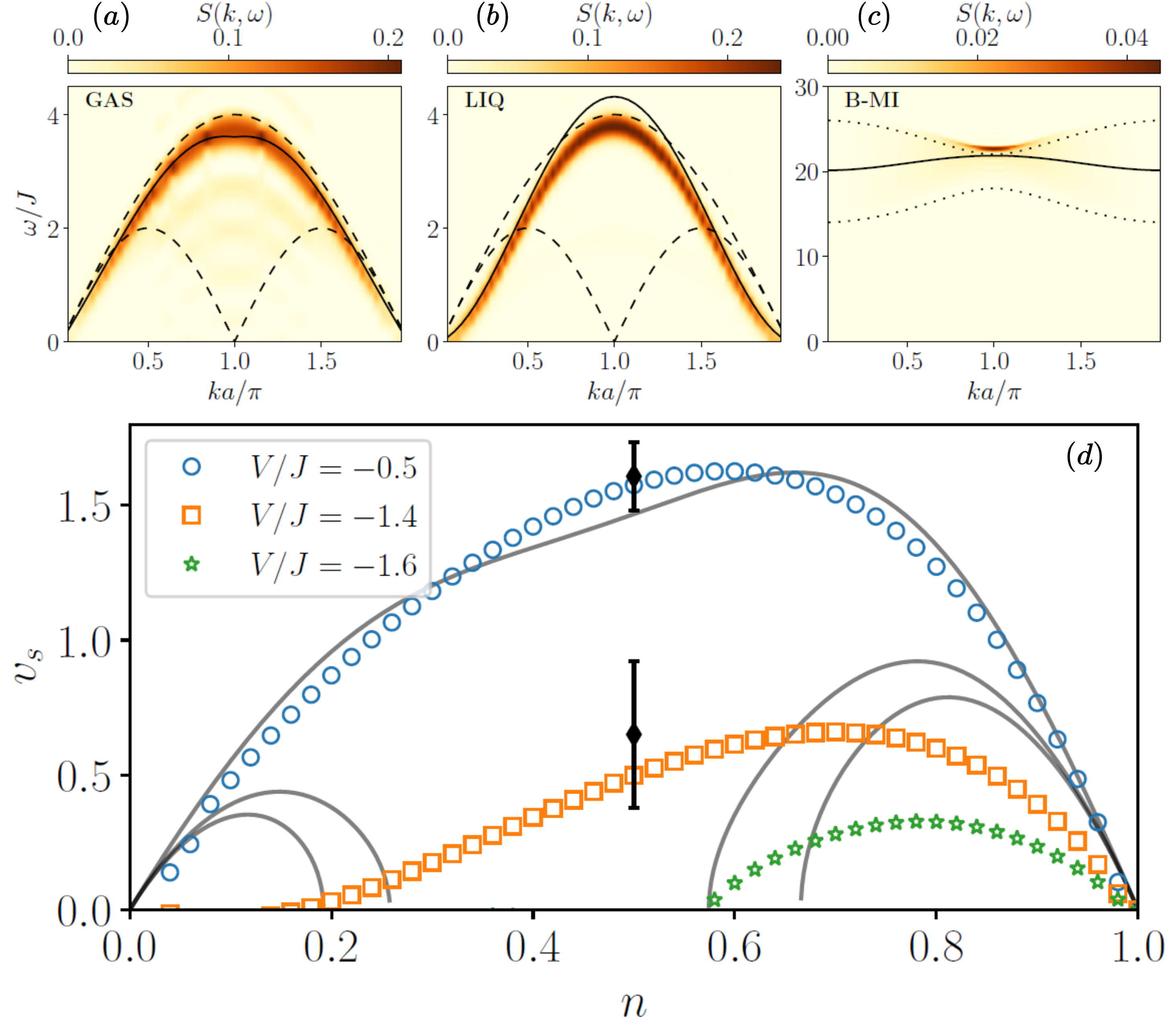}
\caption{(a-c) Dynamic structure factor for (a) the GAS phase ($V/J=-0.5$, $U/J=20$ and $n=25/50$), (b) the LIQ phase ($V/J=-1.4$, $U/J=20$ and $n=25/50$) and (c) the B-MI phase ($V/J=-1.4$, $U/J=20$ and $n=50/50$). 
Upper and lower bounds to the excitation energies for $V/J=0$  are obtained from Bethe ansatz~\cite{Cloizeaux1962,Yamada1969} (dashed lines) for GAS and LIQ and from first-order perturbation theory~\cite{Iucci2006,Tokuno2011,Ejima2012,Ejima2012b} (dotted lines) for B-MI .
(d) Speed of sound as a function of the density for different values of $V/J$ and a fixed $U/J=20$. Symbols from lattice Feynman relation, see the main text;
lines from perturbative results; black diamonds extracted from the dynamic structure factor.
}
\label{Fig:SpeedSound}
\end{figure}
As non-equilibrium properties can be experimentally measured~\cite{Landig2015,Bloch2012}, we investigate the dynamic structure factor $S(k,\omega)$. 
It quantifies the structure and strength of the two-particle excitations allowing an additional verification of the phases.
In Figs.~\ref{Fig:SpeedSound}(a-c), we provide characteristic examples of $S(k,\omega)$ in  different phases. 
The structure of the excitations differs drastically whether the system is compressible or not. 
A single mode exhausts the spectrum in the GAS and LIQ phases. Indeed, the position of the peak in $S(k,\omega)$ is close to the prediction of the Feynman relation, $\omega(k) = \frac{\langle\hat{b}_i^{\dagger}\hat{b}_{i+1} \rangle \epsilon(k)}{nS(k)}$,
derived in the single-mode approximation. Such behavior profoundly varies from that of a
spinon-like spectrum expected at half-filling~\cite{Cloizeaux1962,Yamada1969}, in which modes are populated from the lower (one-particle or -hole excitation) up to upper (two-particle excitation) branches,
see dashed lines in Figs.~\ref{Fig:SpeedSound}(a-b).
The presence of dipolar interactions strongly affects the excitations, creating a dominant mode located close to the upper branch. 
Such a behavior is typical of systems that are softer than the TG gas and possess larger values of the Luttinger parameter, $K>1$~\cite{CauxCalabrese2006}.
We verify that the lowest-energy modes in compressible phases are linear phonons, validating the applicability of the Luttinger Liquid theory.
By employing a flat-top model, i.e., $S(k,\omega)=const$, $\omega_-(k)<\omega<\omega_+(k)$ (see SM for details), we demonstrate that the spectral weight diverges $S(k,v_s k)\propto 1/k^2$ as $k\to 0$, that is the phonon mode is greatly populated, in agreement with the numerical simulations.
The structure of excitations differs dramatically in the B-MI phase, wherein a gap $\Delta$ opens. 
From the flat-top model analysis, we infer that the spectral weight vanishes at small momenta, $S(k,\Delta) \propto k^2$ for $k\to 0$. Instead, the edge of the Brillouin zone, $k_{BZ}=\pi/a$, gets strongly populated, and a sharp peak is formed in $S(k_{BZ},\omega)$ so that the lattice Feynman relation captures well the value of the gap. 
In absence of dipolar interactions, the upper and lower bounds are no longer given by one- or two-particle excitations but rather by doublon-holon excitations, shown with dotted lines.

Having studied the dynamic structure factor in the three phases, we focus now on the speed of sound. The presence of a lattice has strong consequences on sound propagation~\cite{Cazalilla2004_1,Cazalilla2004_2} and other transport properties~\cite{Anderson2019}, as can be traced to the loss of Galilean invariance in the lattice. This produces
a non-trivial dependence of the sound on the density. Since we have shown the lattice Feynman relation correctly captures the sound velocity in GAS/LIQ phases, we will use it in the following.
To obtain the speed of sound $v_s$, we calculate static structure factor $S(k)$, compute the compressibility $\kappa_s$ from the EoS, and employ the non-Galilean version of the Feynman relation~\cite{KRUTITSKY2016,Suppl}, $S(k) = \kappa_s v_s nk/2$ applicable for small momenta. In Fig.~\ref{Fig:SpeedSound}d, we present the sound velocity as a function of the density for different values of the long-range coupling. For weak dipolar attraction, the speed of sound decreases above half-filling and reaches zero at unit filling, signaling the transition to a MI state. As the dipolar strength increases, a liquid state forms, and the sound velocity also vanishes at spinodal density $n_s\leq 1$. Zero value of $v_s$ signals that the homogeneous solution becomes unstable. Finally, when approaching the B-MI phase, the sound velocity nears zero for any value of density showing no stable homogeneous solution exists for $|V|>|V_{\textrm{B-MI}}|$. 

For an analytical estimation of the speed of sound, we employ the non-Galilean invariant Luttinger liquid theory~\cite{Cazalilla2004_1,Cazalilla2004_2}. Within it, the sound velocity $v_s=\sqrt{v_N v_J}$ is determined through the phase $v_J$ and density $v_N$ stiffnesses. Assuming the TG state, we explicitly compute the response of the system and the speed of sound~\cite{Suppl}, see Fig.~\ref{Fig:SpeedSound}b.
For weak dipolar attraction, we note an agreement between the perturbative theory and exact numerical analysis. For the increased strength of attraction, the analytical approach predicts the appearance of a spinodal point, albeit at an incorrect density value. Additionally, we report a qualitative difference at small densities where perturbative theory predicts a finite value of the speed of sound, in stark contrast with numerical results where no stable homogeneous solution exists for $n\leq n_s$. We associate this discrepancy with the formation of molecules at small densities, which is overlooked in the perturbative description.

\textbf{Discussion and outlook.}
Our work shows an unconventional mechanism for liquid formation in strongly correlated lattice systems. Specifically, liquefaction arises due to an interplay between non-local attraction and the superexchange processes originating from short-range repulsion. Our work presents a non-trivial extension of the quantum van der Waals theory to lattice systems. Our predictions apply to lattice systems described by one-dimensional extended Bose-Hubbard Hamiltonians. These could be realized experimentally in different ultracold atomic platforms like dipolar bosons~\cite{Kao296,natale2022strongly}, Rydberg atoms~\cite{labuhn2016tunable,browaeys2020manybody} or even excitonic systems~\cite{lagoin2022checkerboard}. Recently, one-dimensional dipolar bosonic systems were produced experimentally~\cite{Kao296} and loaded into corresponding optical lattices~\cite{natale2022strongly}.

Parallelly, these lattice liquids can be a platform for studying the intriguing properties of non-Galilean invariant liquids with long-range interactions and, for example, their implications on superfluidity. Potential connections between the investigated liquid phases and the magnetic orders expected in the underlying effective spin model also warrant further exploration. Finally, our theory could also be applied to fermionic systems with density-correlated hopping and  long-range attraction. 

\begin{acknowledgments}
\textbf{Acknowledgements.} This work has been funded by Grants No. PID2020-114626GB-I00 and PID2020-113565GB-C21 from the MICIN/AEI/10.13039/501100011033 and by the Ministerio de Economia, Industria y Competitividad (MINECO, Spain) under grants No.~FIS2017-84114-C2-1-P and No.~FIS2017-
87534-P.  We acknowledge financial support from Secretaria d’Universitats i Recerca del Departament
d’Empresa i Coneixement de la Generalitat de
Catalunya, co-funded by the European Union Regional
Development Fund within the ERDF Operational Program of Catalunya (project QuantumCat, ref. 001-P001644). 
R.~O. acknowledges support by the Max Planck Society and the Deutsche Forschungsgemeinschaft (DFG, German Research Foundation) under Germany’s Excellence Strategy – EXC-2111 – 390814868.
\end{acknowledgments}

\bibliographystyle{apsrev4-2}
\bibliography{paperbib}

\begin{thebibliography}{66}%
\makeatletter
\providecommand \@ifxundefined [1]{%
 \@ifx{#1\undefined}
}%
\providecommand \@ifnum [1]{%
 \ifnum #1\expandafter \@firstoftwo
 \else \expandafter \@secondoftwo
 \fi
}%
\providecommand \@ifx [1]{%
 \ifx #1\expandafter \@firstoftwo
 \else \expandafter \@secondoftwo
 \fi
}%
\providecommand \natexlab [1]{#1}%
\providecommand \enquote  [1]{``#1''}%
\providecommand \bibnamefont  [1]{#1}%
\providecommand \bibfnamefont [1]{#1}%
\providecommand \citenamefont [1]{#1}%
\providecommand \href@noop [0]{\@secondoftwo}%
\providecommand \href [0]{\begingroup \@sanitize@url \@href}%
\providecommand \@href[1]{\@@startlink{#1}\@@href}%
\providecommand \@@href[1]{\endgroup#1\@@endlink}%
\providecommand \@sanitize@url [0]{\catcode `\\12\catcode `\$12\catcode
  `\&12\catcode `\#12\catcode `\^12\catcode `\_12\catcode `\%12\relax}%
\providecommand \@@startlink[1]{}%
\providecommand \@@endlink[0]{}%
\providecommand \url  [0]{\begingroup\@sanitize@url \@url }%
\providecommand \@url [1]{\endgroup\@href {#1}{\urlprefix }}%
\providecommand \urlprefix  [0]{URL }%
\providecommand \Eprint [0]{\href }%
\providecommand \doibase [0]{https://doi.org/}%
\providecommand \selectlanguage [0]{\@gobble}%
\providecommand \bibinfo  [0]{\@secondoftwo}%
\providecommand \bibfield  [0]{\@secondoftwo}%
\providecommand \translation [1]{[#1]}%
\providecommand \BibitemOpen [0]{}%
\providecommand \bibitemStop [0]{}%
\providecommand \bibitemNoStop [0]{.\EOS\space}%
\providecommand \EOS [0]{\spacefactor3000\relax}%
\providecommand \BibitemShut  [1]{\csname bibitem#1\endcsname}%
\let\auto@bib@innerbib\@empty
\bibitem [{\citenamefont {Lewenstein}\ \emph {et~al.}(2012)\citenamefont
  {Lewenstein}, \citenamefont {Sanpera},\ and\ \citenamefont
  {Ahufinger}}]{lewenstein2012ultracold}%
  \BibitemOpen
  \bibfield  {author} {\bibinfo {author} {\bibfnamefont {M.}~\bibnamefont
  {Lewenstein}}, \bibinfo {author} {\bibfnamefont {A.}~\bibnamefont
  {Sanpera}},\ and\ \bibinfo {author} {\bibfnamefont {V.}~\bibnamefont
  {Ahufinger}},\ }\href
  {https://doi.org/10.1093/acprof:oso/9780199573127.001.0001} {\emph {\bibinfo
  {title} {Ultracold Atoms in Optical Lattices}}}\ (\bibinfo  {publisher}
  {Oxford University Press},\ \bibinfo {year} {2012})\BibitemShut {NoStop}%
\bibitem [{\citenamefont {Duan}\ \emph
  {et~al.}(2003{\natexlab{a}})\citenamefont {Duan}, \citenamefont {Demler},\
  and\ \citenamefont {Lukin}}]{Duan2003Spin}%
  \BibitemOpen
  \bibfield  {author} {\bibinfo {author} {\bibfnamefont {L.-M.}\ \bibnamefont
  {Duan}}, \bibinfo {author} {\bibfnamefont {E.}~\bibnamefont {Demler}},\ and\
  \bibinfo {author} {\bibfnamefont {M.~D.}\ \bibnamefont {Lukin}},\ }\href
  {https://doi.org/10.1103/PhysRevLett.91.090402} {\bibfield  {journal}
  {\bibinfo  {journal} {Phys. Rev. Lett.}\ }\textbf {\bibinfo {volume} {91}},\
  \bibinfo {pages} {090402} (\bibinfo {year} {2003}{\natexlab{a}})}\BibitemShut
  {NoStop}%
\bibitem [{\citenamefont {Trotzky}\ \emph {et~al.}(2008)\citenamefont
  {Trotzky}, \citenamefont {Cheinet}, \citenamefont {Fölling}, \citenamefont
  {Feld}, \citenamefont {Schnorrberger}, \citenamefont {Rey}, \citenamefont
  {Polkovnikov}, \citenamefont {Demler}, \citenamefont {Lukin},\ and\
  \citenamefont {Bloch}}]{Bloch2008Spin}%
  \BibitemOpen
  \bibfield  {author} {\bibinfo {author} {\bibfnamefont {S.}~\bibnamefont
  {Trotzky}}, \bibinfo {author} {\bibfnamefont {P.}~\bibnamefont {Cheinet}},
  \bibinfo {author} {\bibfnamefont {S.}~\bibnamefont {Fölling}}, \bibinfo
  {author} {\bibfnamefont {M.}~\bibnamefont {Feld}}, \bibinfo {author}
  {\bibfnamefont {U.}~\bibnamefont {Schnorrberger}}, \bibinfo {author}
  {\bibfnamefont {A.~M.}\ \bibnamefont {Rey}}, \bibinfo {author} {\bibfnamefont
  {A.}~\bibnamefont {Polkovnikov}}, \bibinfo {author} {\bibfnamefont {E.~A.}\
  \bibnamefont {Demler}}, \bibinfo {author} {\bibfnamefont {M.~D.}\
  \bibnamefont {Lukin}},\ and\ \bibinfo {author} {\bibfnamefont
  {I.}~\bibnamefont {Bloch}},\ }\href {https://doi.org/10.1126/science.1150841}
  {\bibfield  {journal} {\bibinfo  {journal} {Science}\ }\textbf {\bibinfo
  {volume} {319}},\ \bibinfo {pages} {295} (\bibinfo {year}
  {2008})}\BibitemShut {NoStop}%
\bibitem [{\citenamefont {Duan}\ \emph
  {et~al.}(2003{\natexlab{b}})\citenamefont {Duan}, \citenamefont {Demler},\
  and\ \citenamefont {Lukin}}]{Bloch2012Spin}%
  \BibitemOpen
  \bibfield  {author} {\bibinfo {author} {\bibfnamefont {L.-M.}\ \bibnamefont
  {Duan}}, \bibinfo {author} {\bibfnamefont {E.}~\bibnamefont {Demler}},\ and\
  \bibinfo {author} {\bibfnamefont {M.~D.}\ \bibnamefont {Lukin}},\ }\href
  {https://doi.org/10.1103/PhysRevLett.91.090402} {\bibfield  {journal}
  {\bibinfo  {journal} {Phys. Rev. Lett.}\ }\textbf {\bibinfo {volume} {91}},\
  \bibinfo {pages} {090402} (\bibinfo {year} {2003}{\natexlab{b}})}\BibitemShut
  {NoStop}%
\bibitem [{\citenamefont {Fukuhara}\ \emph {et~al.}(2013)\citenamefont
  {Fukuhara}, \citenamefont {Kantian}, \citenamefont {Endres}, \citenamefont
  {Cheneau}, \citenamefont {Schau{\ss}}, \citenamefont {Hild}, \citenamefont
  {Bellem}, \citenamefont {Schollw\"{o}ck}, \citenamefont {Giamarchi},
  \citenamefont {Gross}, \citenamefont {Bloch},\ and\ \citenamefont
  {Kuhr}}]{Bloch2013Spin}%
  \BibitemOpen
  \bibfield  {author} {\bibinfo {author} {\bibfnamefont {T.}~\bibnamefont
  {Fukuhara}}, \bibinfo {author} {\bibfnamefont {A.}~\bibnamefont {Kantian}},
  \bibinfo {author} {\bibfnamefont {M.}~\bibnamefont {Endres}}, \bibinfo
  {author} {\bibfnamefont {M.}~\bibnamefont {Cheneau}}, \bibinfo {author}
  {\bibfnamefont {P.}~\bibnamefont {Schau{\ss}}}, \bibinfo {author}
  {\bibfnamefont {S.}~\bibnamefont {Hild}}, \bibinfo {author} {\bibfnamefont
  {D.}~\bibnamefont {Bellem}}, \bibinfo {author} {\bibfnamefont
  {U.}~\bibnamefont {Schollw\"{o}ck}}, \bibinfo {author} {\bibfnamefont
  {T.}~\bibnamefont {Giamarchi}}, \bibinfo {author} {\bibfnamefont
  {C.}~\bibnamefont {Gross}}, \bibinfo {author} {\bibfnamefont
  {I.}~\bibnamefont {Bloch}},\ and\ \bibinfo {author} {\bibfnamefont
  {S.}~\bibnamefont {Kuhr}},\ }\href {https://doi.org/10.1038/nphys2561}
  {\bibfield  {journal} {\bibinfo  {journal} {Nature Phys}\ }\textbf {\bibinfo
  {volume} {9}},\ \bibinfo {pages} {235} (\bibinfo {year} {2013})}\BibitemShut
  {NoStop}%
\bibitem [{\citenamefont {Jepsen}\ \emph {et~al.}(2020)\citenamefont {Jepsen},
  \citenamefont {Amato-Grill}, \citenamefont {Dimitrova}, \citenamefont {Ho},
  \citenamefont {Demler},\ and\ \citenamefont {Ketterle}}]{Ketterle2020Spin}%
  \BibitemOpen
  \bibfield  {author} {\bibinfo {author} {\bibfnamefont {P.~N.}\ \bibnamefont
  {Jepsen}}, \bibinfo {author} {\bibfnamefont {J.}~\bibnamefont {Amato-Grill}},
  \bibinfo {author} {\bibfnamefont {I.}~\bibnamefont {Dimitrova}}, \bibinfo
  {author} {\bibfnamefont {W.~W.}\ \bibnamefont {Ho}}, \bibinfo {author}
  {\bibfnamefont {E.}~\bibnamefont {Demler}},\ and\ \bibinfo {author}
  {\bibfnamefont {W.}~\bibnamefont {Ketterle}},\ }\href
  {https://doi.org/10.1038/s41586-020-3033-y} {\bibfield  {journal} {\bibinfo
  {journal} {Nature}\ }\textbf {\bibinfo {volume} {588}},\ \bibinfo {pages}
  {403} (\bibinfo {year} {2020})}\BibitemShut {NoStop}%
\bibitem [{\citenamefont {Baier}\ \emph {et~al.}(2016)\citenamefont {Baier},
  \citenamefont {Mark}, \citenamefont {Petter}, \citenamefont {Aikawa},
  \citenamefont {Chomaz}, \citenamefont {Cai}, \citenamefont {Baranov},
  \citenamefont {Zoller},\ and\ \citenamefont {Ferlaino}}]{baier2016extended}%
  \BibitemOpen
  \bibfield  {author} {\bibinfo {author} {\bibfnamefont {S.}~\bibnamefont
  {Baier}}, \bibinfo {author} {\bibfnamefont {M.~J.}\ \bibnamefont {Mark}},
  \bibinfo {author} {\bibfnamefont {D.}~\bibnamefont {Petter}}, \bibinfo
  {author} {\bibfnamefont {K.}~\bibnamefont {Aikawa}}, \bibinfo {author}
  {\bibfnamefont {L.}~\bibnamefont {Chomaz}}, \bibinfo {author} {\bibfnamefont
  {Z.}~\bibnamefont {Cai}}, \bibinfo {author} {\bibfnamefont {M.}~\bibnamefont
  {Baranov}}, \bibinfo {author} {\bibfnamefont {P.}~\bibnamefont {Zoller}},\
  and\ \bibinfo {author} {\bibfnamefont {F.}~\bibnamefont {Ferlaino}},\ }\href
  {https://doi.org/10.1126/science.aac9812} {\bibfield  {journal} {\bibinfo
  {journal} {Science}\ }\textbf {\bibinfo {volume} {352}},\ \bibinfo {pages}
  {201} (\bibinfo {year} {2016})}\BibitemShut {NoStop}%
\bibitem [{\citenamefont {Natale}\ \emph {et~al.}(2022)\citenamefont {Natale},
  \citenamefont {Bland}, \citenamefont {Gschwendtner}, \citenamefont
  {Lafforgue}, \citenamefont {Gr\"{u}n}, \citenamefont {Patscheider},
  \citenamefont {Mark},\ and\ \citenamefont {Ferlaino}}]{natale2022strongly}%
  \BibitemOpen
  \bibfield  {author} {\bibinfo {author} {\bibfnamefont {G.}~\bibnamefont
  {Natale}}, \bibinfo {author} {\bibfnamefont {T.}~\bibnamefont {Bland}},
  \bibinfo {author} {\bibfnamefont {S.}~\bibnamefont {Gschwendtner}}, \bibinfo
  {author} {\bibfnamefont {L.}~\bibnamefont {Lafforgue}}, \bibinfo {author}
  {\bibfnamefont {D.~S.}\ \bibnamefont {Gr\"{u}n}}, \bibinfo {author}
  {\bibfnamefont {A.}~\bibnamefont {Patscheider}}, \bibinfo {author}
  {\bibfnamefont {M.~J.}\ \bibnamefont {Mark}},\ and\ \bibinfo {author}
  {\bibfnamefont {F.}~\bibnamefont {Ferlaino}},\ }\href
  {https://doi.org/10.48550/ARXIV.2205.03280} {\bibfield  {journal} {\bibinfo
  {journal} {arXiv preprint arXiv:2205.03280}\ ,\ } (\bibinfo {year}
  {2022})}\BibitemShut {NoStop}%
\bibitem [{\citenamefont {Barbiero}\ \emph {et~al.}(2015)\citenamefont
  {Barbiero}, \citenamefont {Menotti}, \citenamefont {Recati},\ and\
  \citenamefont {Santos}}]{Barbiero2015}%
  \BibitemOpen
  \bibfield  {author} {\bibinfo {author} {\bibfnamefont {L.}~\bibnamefont
  {Barbiero}}, \bibinfo {author} {\bibfnamefont {C.}~\bibnamefont {Menotti}},
  \bibinfo {author} {\bibfnamefont {A.}~\bibnamefont {Recati}},\ and\ \bibinfo
  {author} {\bibfnamefont {L.}~\bibnamefont {Santos}},\ }\href
  {https://doi.org/10.1103/PhysRevB.92.180406} {\bibfield  {journal} {\bibinfo
  {journal} {Phys. Rev. B}\ }\textbf {\bibinfo {volume} {92}},\ \bibinfo
  {pages} {180406} (\bibinfo {year} {2015})}\BibitemShut {NoStop}%
\bibitem [{\citenamefont {Li}\ \emph {et~al.}(2020)\citenamefont {Li},
  \citenamefont {Dhar}, \citenamefont {Deng}, \citenamefont {Kasamatsu},
  \citenamefont {Barbiero},\ and\ \citenamefont {Santos}}]{Li2020}%
  \BibitemOpen
  \bibfield  {author} {\bibinfo {author} {\bibfnamefont {W.}~\bibnamefont
  {Li}}, \bibinfo {author} {\bibfnamefont {A.}~\bibnamefont {Dhar}}, \bibinfo
  {author} {\bibfnamefont {X.}~\bibnamefont {Deng}}, \bibinfo {author}
  {\bibfnamefont {K.}~\bibnamefont {Kasamatsu}}, \bibinfo {author}
  {\bibfnamefont {L.}~\bibnamefont {Barbiero}},\ and\ \bibinfo {author}
  {\bibfnamefont {L.}~\bibnamefont {Santos}},\ }\href
  {https://doi.org/10.1103/PhysRevLett.124.010404} {\bibfield  {journal}
  {\bibinfo  {journal} {Phys. Rev. Lett.}\ }\textbf {\bibinfo {volume} {124}},\
  \bibinfo {pages} {010404} (\bibinfo {year} {2020})}\BibitemShut {NoStop}%
\bibitem [{\citenamefont {Kraus}\ \emph {et~al.}(2022)\citenamefont {Kraus},
  \citenamefont {Chanda}, \citenamefont {Zakrzewski},\ and\ \citenamefont
  {Morigi}}]{kraus2021quantum}%
  \BibitemOpen
  \bibfield  {author} {\bibinfo {author} {\bibfnamefont {R.}~\bibnamefont
  {Kraus}}, \bibinfo {author} {\bibfnamefont {T.}~\bibnamefont {Chanda}},
  \bibinfo {author} {\bibfnamefont {J.}~\bibnamefont {Zakrzewski}},\ and\
  \bibinfo {author} {\bibfnamefont {G.}~\bibnamefont {Morigi}},\ }\href
  {https://doi.org/10.1088/1367-2630/aaf295} {\bibfield  {journal} {\bibinfo
  {journal} {Phys. Rev. B}\ }\textbf {\bibinfo {volume} {106}},\ \bibinfo
  {pages} {035144} (\bibinfo {year} {2022})}\BibitemShut {NoStop}%
\bibitem [{\citenamefont {Fraxanet}\ \emph {et~al.}(2022)\citenamefont
  {Fraxanet}, \citenamefont {Gonz\'alez-Cuadra}, \citenamefont {Pfau},
  \citenamefont {Lewenstein}, \citenamefont {Langen},\ and\ \citenamefont
  {Barbiero}}]{Barbiero2022}%
  \BibitemOpen
  \bibfield  {author} {\bibinfo {author} {\bibfnamefont {J.}~\bibnamefont
  {Fraxanet}}, \bibinfo {author} {\bibfnamefont {D.}~\bibnamefont
  {Gonz\'alez-Cuadra}}, \bibinfo {author} {\bibfnamefont {T.}~\bibnamefont
  {Pfau}}, \bibinfo {author} {\bibfnamefont {M.}~\bibnamefont {Lewenstein}},
  \bibinfo {author} {\bibfnamefont {T.}~\bibnamefont {Langen}},\ and\ \bibinfo
  {author} {\bibfnamefont {L.}~\bibnamefont {Barbiero}},\ }\href
  {https://doi.org/10.1103/PhysRevLett.128.043402} {\bibfield  {journal}
  {\bibinfo  {journal} {Phys. Rev. Lett.}\ }\textbf {\bibinfo {volume} {128}},\
  \bibinfo {pages} {043402} (\bibinfo {year} {2022})}\BibitemShut {NoStop}%
\bibitem [{\citenamefont {Van Der~Waals}\ and\ \citenamefont
  {Rowlinson}(2004)}]{van2004continuity}%
  \BibitemOpen
  \bibfield  {author} {\bibinfo {author} {\bibfnamefont {J.~D.}\ \bibnamefont
  {Van Der~Waals}}\ and\ \bibinfo {author} {\bibfnamefont {J.~S.}\ \bibnamefont
  {Rowlinson}},\ }\href@noop {} {\emph {\bibinfo {title} {On the continuity of
  the gaseous and liquid states}}}\ (\bibinfo  {publisher} {Courier
  Corporation},\ \bibinfo {year} {2004})\BibitemShut {NoStop}%
\bibitem [{\citenamefont {Schmitt}\ \emph {et~al.}(2016)\citenamefont
  {Schmitt}, \citenamefont {Wenzel}, \citenamefont {B\"{o}ttcher},
  \citenamefont {Ferrier-Barbut},\ and\ \citenamefont {Pfau}}]{Schmitt2016}%
  \BibitemOpen
  \bibfield  {author} {\bibinfo {author} {\bibfnamefont {M.}~\bibnamefont
  {Schmitt}}, \bibinfo {author} {\bibfnamefont {M.}~\bibnamefont {Wenzel}},
  \bibinfo {author} {\bibfnamefont {F.}~\bibnamefont {B\"{o}ttcher}}, \bibinfo
  {author} {\bibfnamefont {I.}~\bibnamefont {Ferrier-Barbut}},\ and\ \bibinfo
  {author} {\bibfnamefont {T.}~\bibnamefont {Pfau}},\ }\href
  {https://doi.org/10.1038/nature20126} {\bibfield  {journal} {\bibinfo
  {journal} {Nature}\ }\textbf {\bibinfo {volume} {539}},\ \bibinfo {pages}
  {259} (\bibinfo {year} {2016})}\BibitemShut {NoStop}%
\bibitem [{\citenamefont {Ferrier-Barbut}\ \emph {et~al.}(2016)\citenamefont
  {Ferrier-Barbut}, \citenamefont {Kadau}, \citenamefont {Schmitt},
  \citenamefont {Wenzel},\ and\ \citenamefont {Pfau}}]{Barbut2016}%
  \BibitemOpen
  \bibfield  {author} {\bibinfo {author} {\bibfnamefont {I.}~\bibnamefont
  {Ferrier-Barbut}}, \bibinfo {author} {\bibfnamefont {H.}~\bibnamefont
  {Kadau}}, \bibinfo {author} {\bibfnamefont {M.}~\bibnamefont {Schmitt}},
  \bibinfo {author} {\bibfnamefont {M.}~\bibnamefont {Wenzel}},\ and\ \bibinfo
  {author} {\bibfnamefont {T.}~\bibnamefont {Pfau}},\ }\href
  {https://doi.org/10.1103/PhysRevLett.116.215301} {\bibfield  {journal}
  {\bibinfo  {journal} {Phys. Rev. Lett.}\ }\textbf {\bibinfo {volume} {116}},\
  \bibinfo {pages} {215301} (\bibinfo {year} {2016})}\BibitemShut {NoStop}%
\bibitem [{\citenamefont {Chomaz}\ \emph {et~al.}(2016)\citenamefont {Chomaz},
  \citenamefont {Baier}, \citenamefont {Petter}, \citenamefont {Mark},
  \citenamefont {W\"achtler}, \citenamefont {Santos},\ and\ \citenamefont
  {Ferlaino}}]{Chomaz2016}%
  \BibitemOpen
  \bibfield  {author} {\bibinfo {author} {\bibfnamefont {L.}~\bibnamefont
  {Chomaz}}, \bibinfo {author} {\bibfnamefont {S.}~\bibnamefont {Baier}},
  \bibinfo {author} {\bibfnamefont {D.}~\bibnamefont {Petter}}, \bibinfo
  {author} {\bibfnamefont {M.~J.}\ \bibnamefont {Mark}}, \bibinfo {author}
  {\bibfnamefont {F.}~\bibnamefont {W\"achtler}}, \bibinfo {author}
  {\bibfnamefont {L.}~\bibnamefont {Santos}},\ and\ \bibinfo {author}
  {\bibfnamefont {F.}~\bibnamefont {Ferlaino}},\ }\href
  {https://doi.org/10.1103/PhysRevX.6.041039} {\bibfield  {journal} {\bibinfo
  {journal} {Phys. Rev. X}\ }\textbf {\bibinfo {volume} {6}},\ \bibinfo {pages}
  {041039} (\bibinfo {year} {2016})}\BibitemShut {NoStop}%
\bibitem [{\citenamefont {Ferrier-Barbut}\ \emph {et~al.}(2018)\citenamefont
  {Ferrier-Barbut}, \citenamefont {Wenzel}, \citenamefont {B\"ottcher},
  \citenamefont {Langen}, \citenamefont {Isoard}, \citenamefont {Stringari},\
  and\ \citenamefont {Pfau}}]{Barbut2018}%
  \BibitemOpen
  \bibfield  {author} {\bibinfo {author} {\bibfnamefont {I.}~\bibnamefont
  {Ferrier-Barbut}}, \bibinfo {author} {\bibfnamefont {M.}~\bibnamefont
  {Wenzel}}, \bibinfo {author} {\bibfnamefont {F.}~\bibnamefont {B\"ottcher}},
  \bibinfo {author} {\bibfnamefont {T.}~\bibnamefont {Langen}}, \bibinfo
  {author} {\bibfnamefont {M.}~\bibnamefont {Isoard}}, \bibinfo {author}
  {\bibfnamefont {S.}~\bibnamefont {Stringari}},\ and\ \bibinfo {author}
  {\bibfnamefont {T.}~\bibnamefont {Pfau}},\ }\href
  {https://doi.org/10.1103/PhysRevLett.120.160402} {\bibfield  {journal}
  {\bibinfo  {journal} {Phys. Rev. Lett.}\ }\textbf {\bibinfo {volume} {120}},\
  \bibinfo {pages} {160402} (\bibinfo {year} {2018})}\BibitemShut {NoStop}%
\bibitem [{\citenamefont {B\"ottcher}\ \emph {et~al.}(2019)\citenamefont
  {B\"ottcher}, \citenamefont {Wenzel}, \citenamefont {Schmidt}, \citenamefont
  {Guo}, \citenamefont {Langen}, \citenamefont {Ferrier-Barbut}, \citenamefont
  {Pfau}, \citenamefont {Bomb\'{\i}n}, \citenamefont {S\'anchez-Baena},
  \citenamefont {Boronat},\ and\ \citenamefont {Mazzanti}}]{Boettcher2019}%
  \BibitemOpen
  \bibfield  {author} {\bibinfo {author} {\bibfnamefont {F.}~\bibnamefont
  {B\"ottcher}}, \bibinfo {author} {\bibfnamefont {M.}~\bibnamefont {Wenzel}},
  \bibinfo {author} {\bibfnamefont {J.-N.}\ \bibnamefont {Schmidt}}, \bibinfo
  {author} {\bibfnamefont {M.}~\bibnamefont {Guo}}, \bibinfo {author}
  {\bibfnamefont {T.}~\bibnamefont {Langen}}, \bibinfo {author} {\bibfnamefont
  {I.}~\bibnamefont {Ferrier-Barbut}}, \bibinfo {author} {\bibfnamefont
  {T.}~\bibnamefont {Pfau}}, \bibinfo {author} {\bibfnamefont {R.}~\bibnamefont
  {Bomb\'{\i}n}}, \bibinfo {author} {\bibfnamefont {J.}~\bibnamefont
  {S\'anchez-Baena}}, \bibinfo {author} {\bibfnamefont {J.}~\bibnamefont
  {Boronat}},\ and\ \bibinfo {author} {\bibfnamefont {F.}~\bibnamefont
  {Mazzanti}},\ }\href {https://doi.org/10.1103/PhysRevResearch.1.033088}
  {\bibfield  {journal} {\bibinfo  {journal} {Phys. Rev. Research}\ }\textbf
  {\bibinfo {volume} {1}},\ \bibinfo {pages} {033088} (\bibinfo {year}
  {2019})}\BibitemShut {NoStop}%
\bibitem [{\citenamefont {Cabrera}\ \emph {et~al.}(2018)\citenamefont
  {Cabrera}, \citenamefont {Tanzi}, \citenamefont {Sanz}, \citenamefont
  {Naylor}, \citenamefont {Thomas}, \citenamefont {Cheiney},\ and\
  \citenamefont {Tarruell}}]{Tarruell2018}%
  \BibitemOpen
  \bibfield  {author} {\bibinfo {author} {\bibfnamefont {C.~R.}\ \bibnamefont
  {Cabrera}}, \bibinfo {author} {\bibfnamefont {L.}~\bibnamefont {Tanzi}},
  \bibinfo {author} {\bibfnamefont {J.}~\bibnamefont {Sanz}}, \bibinfo {author}
  {\bibfnamefont {B.}~\bibnamefont {Naylor}}, \bibinfo {author} {\bibfnamefont
  {P.}~\bibnamefont {Thomas}}, \bibinfo {author} {\bibfnamefont
  {P.}~\bibnamefont {Cheiney}},\ and\ \bibinfo {author} {\bibfnamefont
  {L.}~\bibnamefont {Tarruell}},\ }\href
  {https://doi.org/10.1126/science.aao5686} {\bibfield  {journal} {\bibinfo
  {journal} {Science}\ }\textbf {\bibinfo {volume} {359}},\ \bibinfo {pages}
  {301} (\bibinfo {year} {2018})}\BibitemShut {NoStop}%
\bibitem [{\citenamefont {Cheiney}\ \emph {et~al.}(2018)\citenamefont
  {Cheiney}, \citenamefont {Cabrera}, \citenamefont {Sanz}, \citenamefont
  {Naylor}, \citenamefont {Tanzi},\ and\ \citenamefont
  {Tarruell}}]{Tarruell2018a}%
  \BibitemOpen
  \bibfield  {author} {\bibinfo {author} {\bibfnamefont {P.}~\bibnamefont
  {Cheiney}}, \bibinfo {author} {\bibfnamefont {C.~R.}\ \bibnamefont
  {Cabrera}}, \bibinfo {author} {\bibfnamefont {J.}~\bibnamefont {Sanz}},
  \bibinfo {author} {\bibfnamefont {B.}~\bibnamefont {Naylor}}, \bibinfo
  {author} {\bibfnamefont {L.}~\bibnamefont {Tanzi}},\ and\ \bibinfo {author}
  {\bibfnamefont {L.}~\bibnamefont {Tarruell}},\ }\href
  {https://doi.org/10.1103/PhysRevLett.120.135301} {\bibfield  {journal}
  {\bibinfo  {journal} {Phys. Rev. Lett.}\ }\textbf {\bibinfo {volume} {120}},\
  \bibinfo {pages} {135301} (\bibinfo {year} {2018})}\BibitemShut {NoStop}%
\bibitem [{\citenamefont {Semeghini}\ \emph {et~al.}(2018)\citenamefont
  {Semeghini}, \citenamefont {Ferioli}, \citenamefont {Masi}, \citenamefont
  {Mazzinghi}, \citenamefont {Wolswijk}, \citenamefont {Minardi}, \citenamefont
  {Modugno}, \citenamefont {Modugno}, \citenamefont {Inguscio},\ and\
  \citenamefont {Fattori}}]{Fattori2018}%
  \BibitemOpen
  \bibfield  {author} {\bibinfo {author} {\bibfnamefont {G.}~\bibnamefont
  {Semeghini}}, \bibinfo {author} {\bibfnamefont {G.}~\bibnamefont {Ferioli}},
  \bibinfo {author} {\bibfnamefont {L.}~\bibnamefont {Masi}}, \bibinfo {author}
  {\bibfnamefont {C.}~\bibnamefont {Mazzinghi}}, \bibinfo {author}
  {\bibfnamefont {L.}~\bibnamefont {Wolswijk}}, \bibinfo {author}
  {\bibfnamefont {F.}~\bibnamefont {Minardi}}, \bibinfo {author} {\bibfnamefont
  {M.}~\bibnamefont {Modugno}}, \bibinfo {author} {\bibfnamefont
  {G.}~\bibnamefont {Modugno}}, \bibinfo {author} {\bibfnamefont
  {M.}~\bibnamefont {Inguscio}},\ and\ \bibinfo {author} {\bibfnamefont
  {M.}~\bibnamefont {Fattori}},\ }\href
  {https://doi.org/10.1103/PhysRevLett.120.235301} {\bibfield  {journal}
  {\bibinfo  {journal} {Phys. Rev. Lett.}\ }\textbf {\bibinfo {volume} {120}},\
  \bibinfo {pages} {235301} (\bibinfo {year} {2018})}\BibitemShut {NoStop}%
\bibitem [{\citenamefont {Petrov}(2015)}]{Petrov2015}%
  \BibitemOpen
  \bibfield  {author} {\bibinfo {author} {\bibfnamefont {D.~S.}\ \bibnamefont
  {Petrov}},\ }\href {https://doi.org/10.1103/PhysRevLett.115.155302}
  {\bibfield  {journal} {\bibinfo  {journal} {Phys. Rev. Lett.}\ }\textbf
  {\bibinfo {volume} {115}},\ \bibinfo {pages} {155302} (\bibinfo {year}
  {2015})}\BibitemShut {NoStop}%
\bibitem [{\citenamefont {Lee}\ \emph {et~al.}(1957)\citenamefont {Lee},
  \citenamefont {Huang},\ and\ \citenamefont {Yang}}]{LHY1957}%
  \BibitemOpen
  \bibfield  {author} {\bibinfo {author} {\bibfnamefont {T.~D.}\ \bibnamefont
  {Lee}}, \bibinfo {author} {\bibfnamefont {K.}~\bibnamefont {Huang}},\ and\
  \bibinfo {author} {\bibfnamefont {C.~N.}\ \bibnamefont {Yang}},\ }\href
  {https://doi.org/10.1103/PhysRev.106.1135} {\bibfield  {journal} {\bibinfo
  {journal} {Phys. Rev.}\ }\textbf {\bibinfo {volume} {106}},\ \bibinfo {pages}
  {1135} (\bibinfo {year} {1957})}\BibitemShut {NoStop}%
\bibitem [{\citenamefont {Lima}\ and\ \citenamefont
  {Pelster}(2011)}]{Lima2011}%
  \BibitemOpen
  \bibfield  {author} {\bibinfo {author} {\bibfnamefont {A.~R.~P.}\
  \bibnamefont {Lima}}\ and\ \bibinfo {author} {\bibfnamefont {A.}~\bibnamefont
  {Pelster}},\ }\href {https://doi.org/10.1103/PhysRevA.84.041604} {\bibfield
  {journal} {\bibinfo  {journal} {Phys. Rev. A}\ }\textbf {\bibinfo {volume}
  {84}},\ \bibinfo {pages} {041604} (\bibinfo {year} {2011})}\BibitemShut
  {NoStop}%
\bibitem [{\citenamefont {Lima}\ and\ \citenamefont
  {Pelster}(2012)}]{Lima2012}%
  \BibitemOpen
  \bibfield  {author} {\bibinfo {author} {\bibfnamefont {A.~R.~P.}\
  \bibnamefont {Lima}}\ and\ \bibinfo {author} {\bibfnamefont {A.}~\bibnamefont
  {Pelster}},\ }\href {https://doi.org/10.1103/PhysRevA.86.063609} {\bibfield
  {journal} {\bibinfo  {journal} {Phys. Rev. A}\ }\textbf {\bibinfo {volume}
  {86}},\ \bibinfo {pages} {063609} (\bibinfo {year} {2012})}\BibitemShut
  {NoStop}%
\bibitem [{\citenamefont {Petrov}\ and\ \citenamefont
  {Astrakharchik}(2016)}]{Petrov2016}%
  \BibitemOpen
  \bibfield  {author} {\bibinfo {author} {\bibfnamefont {D.~S.}\ \bibnamefont
  {Petrov}}\ and\ \bibinfo {author} {\bibfnamefont {G.~E.}\ \bibnamefont
  {Astrakharchik}},\ }\href {https://doi.org/10.1103/PhysRevLett.117.100401}
  {\bibfield  {journal} {\bibinfo  {journal} {Phys. Rev. Lett.}\ }\textbf
  {\bibinfo {volume} {117}},\ \bibinfo {pages} {100401} (\bibinfo {year}
  {2016})}\BibitemShut {NoStop}%
\bibitem [{\citenamefont {Edler}\ \emph {et~al.}(2017)\citenamefont {Edler},
  \citenamefont {Mishra}, \citenamefont {W\"achtler}, \citenamefont {Nath},
  \citenamefont {Sinha},\ and\ \citenamefont {Santos}}]{Santos2017}%
  \BibitemOpen
  \bibfield  {author} {\bibinfo {author} {\bibfnamefont {D.}~\bibnamefont
  {Edler}}, \bibinfo {author} {\bibfnamefont {C.}~\bibnamefont {Mishra}},
  \bibinfo {author} {\bibfnamefont {F.}~\bibnamefont {W\"achtler}}, \bibinfo
  {author} {\bibfnamefont {R.}~\bibnamefont {Nath}}, \bibinfo {author}
  {\bibfnamefont {S.}~\bibnamefont {Sinha}},\ and\ \bibinfo {author}
  {\bibfnamefont {L.}~\bibnamefont {Santos}},\ }\href
  {https://doi.org/10.1103/PhysRevLett.119.050403} {\bibfield  {journal}
  {\bibinfo  {journal} {Phys. Rev. Lett.}\ }\textbf {\bibinfo {volume} {119}},\
  \bibinfo {pages} {050403} (\bibinfo {year} {2017})}\BibitemShut {NoStop}%
\bibitem [{\citenamefont {Astrakharchik}\ and\ \citenamefont
  {Malomed}(2018)}]{Malomed2018}%
  \BibitemOpen
  \bibfield  {author} {\bibinfo {author} {\bibfnamefont {G.~E.}\ \bibnamefont
  {Astrakharchik}}\ and\ \bibinfo {author} {\bibfnamefont {B.~A.}\ \bibnamefont
  {Malomed}},\ }\href {https://doi.org/10.1103/PhysRevA.98.013631} {\bibfield
  {journal} {\bibinfo  {journal} {Phys. Rev. A}\ }\textbf {\bibinfo {volume}
  {98}},\ \bibinfo {pages} {013631} (\bibinfo {year} {2018})}\BibitemShut
  {NoStop}%
\bibitem [{\citenamefont {Zin}\ \emph {et~al.}(2018)\citenamefont {Zin},
  \citenamefont {Pylak}, \citenamefont {Wasak}, \citenamefont {Gajda},\ and\
  \citenamefont {Idziaszek}}]{Zin2018}%
  \BibitemOpen
  \bibfield  {author} {\bibinfo {author} {\bibfnamefont {P.}~\bibnamefont
  {Zin}}, \bibinfo {author} {\bibfnamefont {M.}~\bibnamefont {Pylak}}, \bibinfo
  {author} {\bibfnamefont {T.}~\bibnamefont {Wasak}}, \bibinfo {author}
  {\bibfnamefont {M.}~\bibnamefont {Gajda}},\ and\ \bibinfo {author}
  {\bibfnamefont {Z.}~\bibnamefont {Idziaszek}},\ }\href
  {https://doi.org/10.1103/PhysRevA.98.051603} {\bibfield  {journal} {\bibinfo
  {journal} {Phys. Rev. A}\ }\textbf {\bibinfo {volume} {98}},\ \bibinfo
  {pages} {051603} (\bibinfo {year} {2018})}\BibitemShut {NoStop}%
\bibitem [{\citenamefont {Ilg}\ \emph {et~al.}(2018)\citenamefont {Ilg},
  \citenamefont {Kumlin}, \citenamefont {Santos}, \citenamefont {Petrov},\ and\
  \citenamefont {B\"uchler}}]{Ilg2018}%
  \BibitemOpen
  \bibfield  {author} {\bibinfo {author} {\bibfnamefont {T.}~\bibnamefont
  {Ilg}}, \bibinfo {author} {\bibfnamefont {J.}~\bibnamefont {Kumlin}},
  \bibinfo {author} {\bibfnamefont {L.}~\bibnamefont {Santos}}, \bibinfo
  {author} {\bibfnamefont {D.~S.}\ \bibnamefont {Petrov}},\ and\ \bibinfo
  {author} {\bibfnamefont {H.~P.}\ \bibnamefont {B\"uchler}},\ }\href
  {https://doi.org/10.1103/PhysRevA.98.051604} {\bibfield  {journal} {\bibinfo
  {journal} {Phys. Rev. A}\ }\textbf {\bibinfo {volume} {98}},\ \bibinfo
  {pages} {051604} (\bibinfo {year} {2018})}\BibitemShut {NoStop}%
\bibitem [{\citenamefont {Rakshit}\ \emph {et~al.}(2019)\citenamefont
  {Rakshit}, \citenamefont {Karpiuk}, \citenamefont {Zin}, \citenamefont
  {Brewczyk}, \citenamefont {Lewenstein},\ and\ \citenamefont
  {Gajda}}]{Rakshit2019}%
  \BibitemOpen
  \bibfield  {author} {\bibinfo {author} {\bibfnamefont {D.}~\bibnamefont
  {Rakshit}}, \bibinfo {author} {\bibfnamefont {T.}~\bibnamefont {Karpiuk}},
  \bibinfo {author} {\bibfnamefont {P.}~\bibnamefont {Zin}}, \bibinfo {author}
  {\bibfnamefont {M.}~\bibnamefont {Brewczyk}}, \bibinfo {author}
  {\bibfnamefont {M.}~\bibnamefont {Lewenstein}},\ and\ \bibinfo {author}
  {\bibfnamefont {M.}~\bibnamefont {Gajda}},\ }\href
  {https://doi.org/10.1088/1367-2630/ab2ce3} {\bibfield  {journal} {\bibinfo
  {journal} {New Journal of Physics}\ }\textbf {\bibinfo {volume} {21}},\
  \bibinfo {pages} {073027} (\bibinfo {year} {2019})}\BibitemShut {NoStop}%
\bibitem [{\citenamefont {O\l{}dziejewski}\ \emph {et~al.}(2020)\citenamefont
  {O\l{}dziejewski}, \citenamefont {G\'orecki}, \citenamefont {Paw\l{}owski},\
  and\ \citenamefont {Rz\k{a}\ifmmode~\dot{z}\else
  \.{z}\fi{}ewski}}]{Oldziejewski2020}%
  \BibitemOpen
  \bibfield  {author} {\bibinfo {author} {\bibfnamefont {R.}~\bibnamefont
  {O\l{}dziejewski}}, \bibinfo {author} {\bibfnamefont {W.}~\bibnamefont
  {G\'orecki}}, \bibinfo {author} {\bibfnamefont {K.}~\bibnamefont
  {Paw\l{}owski}},\ and\ \bibinfo {author} {\bibfnamefont {K.}~\bibnamefont
  {Rz\k{a}\ifmmode~\dot{z}\else \.{z}\fi{}ewski}},\ }\href
  {https://doi.org/10.1103/PhysRevLett.124.090401} {\bibfield  {journal}
  {\bibinfo  {journal} {Phys. Rev. Lett.}\ }\textbf {\bibinfo {volume} {124}},\
  \bibinfo {pages} {090401} (\bibinfo {year} {2020})}\BibitemShut {NoStop}%
\bibitem [{\citenamefont {Palo}\ \emph {et~al.}(2022)\citenamefont {Palo},
  \citenamefont {Orignac},\ and\ \citenamefont {Citro}}]{de2022}%
  \BibitemOpen
  \bibfield  {author} {\bibinfo {author} {\bibfnamefont {S.~D.}\ \bibnamefont
  {Palo}}, \bibinfo {author} {\bibfnamefont {E.}~\bibnamefont {Orignac}},\ and\
  \bibinfo {author} {\bibfnamefont {R.}~\bibnamefont {Citro}},\ }\href
  {https://doi.org/10.1103/physrevb.106.014503} {\bibfield  {journal} {\bibinfo
   {journal} {Phys. Rev. B}\ }\textbf {\bibinfo {volume} {106}},\ \bibinfo
  {pages} {014503} (\bibinfo {year} {2022})}\BibitemShut {NoStop}%
\bibitem [{\citenamefont {Guijarro}\ \emph {et~al.}(2022)\citenamefont
  {Guijarro}, \citenamefont {Astrakharchik},\ and\ \citenamefont
  {Boronat}}]{Guijarro2022}%
  \BibitemOpen
  \bibfield  {author} {\bibinfo {author} {\bibfnamefont {G.}~\bibnamefont
  {Guijarro}}, \bibinfo {author} {\bibfnamefont {G.~E.}\ \bibnamefont
  {Astrakharchik}},\ and\ \bibinfo {author} {\bibfnamefont {J.}~\bibnamefont
  {Boronat}},\ }\href {https://doi.org/10.1103/PhysRevLett.128.063401}
  {\bibfield  {journal} {\bibinfo  {journal} {Phys. Rev. Lett.}\ }\textbf
  {\bibinfo {volume} {128}},\ \bibinfo {pages} {063401} (\bibinfo {year}
  {2022})}\BibitemShut {NoStop}%
\bibitem [{\citenamefont {Morera}\ \emph {et~al.}(2020)\citenamefont {Morera},
  \citenamefont {Astrakharchik}, \citenamefont {Polls},\ and\ \citenamefont
  {Juli\'a-D\'{\i}az}}]{Morera2020}%
  \BibitemOpen
  \bibfield  {author} {\bibinfo {author} {\bibfnamefont {I.}~\bibnamefont
  {Morera}}, \bibinfo {author} {\bibfnamefont {G.~E.}\ \bibnamefont
  {Astrakharchik}}, \bibinfo {author} {\bibfnamefont {A.}~\bibnamefont
  {Polls}},\ and\ \bibinfo {author} {\bibfnamefont {B.}~\bibnamefont
  {Juli\'a-D\'{\i}az}},\ }\href
  {https://doi.org/10.1103/PhysRevResearch.2.022008} {\bibfield  {journal}
  {\bibinfo  {journal} {Phys. Rev. Research}\ }\textbf {\bibinfo {volume}
  {2}},\ \bibinfo {pages} {022008} (\bibinfo {year} {2020})}\BibitemShut
  {NoStop}%
\bibitem [{\citenamefont {Morera}\ \emph {et~al.}(2021)\citenamefont {Morera},
  \citenamefont {Astrakharchik}, \citenamefont {Polls},\ and\ \citenamefont
  {Juli\'a-D\'{\i}az}}]{Morera2021}%
  \BibitemOpen
  \bibfield  {author} {\bibinfo {author} {\bibfnamefont {I.}~\bibnamefont
  {Morera}}, \bibinfo {author} {\bibfnamefont {G.~E.}\ \bibnamefont
  {Astrakharchik}}, \bibinfo {author} {\bibfnamefont {A.}~\bibnamefont
  {Polls}},\ and\ \bibinfo {author} {\bibfnamefont {B.}~\bibnamefont
  {Juli\'a-D\'{\i}az}},\ }\href
  {https://doi.org/10.1103/PhysRevLett.126.023001} {\bibfield  {journal}
  {\bibinfo  {journal} {Phys. Rev. Lett.}\ }\textbf {\bibinfo {volume} {126}},\
  \bibinfo {pages} {023001} (\bibinfo {year} {2021})}\BibitemShut {NoStop}%
\bibitem [{\citenamefont {Kao}\ \emph {et~al.}(2021)\citenamefont {Kao},
  \citenamefont {Li}, \citenamefont {Lin}, \citenamefont {Gopalakrishnan},\
  and\ \citenamefont {Lev}}]{Kao296}%
  \BibitemOpen
  \bibfield  {author} {\bibinfo {author} {\bibfnamefont {W.}~\bibnamefont
  {Kao}}, \bibinfo {author} {\bibfnamefont {K.-Y.}\ \bibnamefont {Li}},
  \bibinfo {author} {\bibfnamefont {K.-Y.}\ \bibnamefont {Lin}}, \bibinfo
  {author} {\bibfnamefont {S.}~\bibnamefont {Gopalakrishnan}},\ and\ \bibinfo
  {author} {\bibfnamefont {B.~L.}\ \bibnamefont {Lev}},\ }\href
  {https://doi.org/10.1126/science.abb4928} {\bibfield  {journal} {\bibinfo
  {journal} {Science}\ }\textbf {\bibinfo {volume} {371}},\ \bibinfo {pages}
  {296} (\bibinfo {year} {2021})}\BibitemShut {NoStop}%
\bibitem [{\citenamefont {Labuhn}\ \emph {et~al.}(2016)\citenamefont {Labuhn},
  \citenamefont {Barredo}, \citenamefont {Ravets}, \citenamefont
  {de~Léséleuc}, \citenamefont {Macrì}, \citenamefont {Lahaye},\ and\
  \citenamefont {Browaeys}}]{labuhn2016tunable}%
  \BibitemOpen
  \bibfield  {author} {\bibinfo {author} {\bibfnamefont {H.}~\bibnamefont
  {Labuhn}}, \bibinfo {author} {\bibfnamefont {D.}~\bibnamefont {Barredo}},
  \bibinfo {author} {\bibfnamefont {S.}~\bibnamefont {Ravets}}, \bibinfo
  {author} {\bibfnamefont {S.}~\bibnamefont {de~Léséleuc}}, \bibinfo {author}
  {\bibfnamefont {T.}~\bibnamefont {Macrì}}, \bibinfo {author} {\bibfnamefont
  {T.}~\bibnamefont {Lahaye}},\ and\ \bibinfo {author} {\bibfnamefont
  {A.}~\bibnamefont {Browaeys}},\ }\href {https://doi.org/10.1038/nature18274}
  {\bibfield  {journal} {\bibinfo  {journal} {Nature}\ }\textbf {\bibinfo
  {volume} {534}},\ \bibinfo {pages} {667} (\bibinfo {year}
  {2016})}\BibitemShut {NoStop}%
\bibitem [{\citenamefont {Browaeys}\ and\ \citenamefont
  {Lahaye}(2020)}]{browaeys2020manybody}%
  \BibitemOpen
  \bibfield  {author} {\bibinfo {author} {\bibfnamefont {A.}~\bibnamefont
  {Browaeys}}\ and\ \bibinfo {author} {\bibfnamefont {T.}~\bibnamefont
  {Lahaye}},\ }\href {https://doi.org/10.1038/s41567-019-0733-z} {\bibfield
  {journal} {\bibinfo  {journal} {Nat. Phys.}\ }\textbf {\bibinfo {volume}
  {16}},\ \bibinfo {pages} {132} (\bibinfo {year} {2020})}\BibitemShut
  {NoStop}%
\bibitem [{\citenamefont {Lagoin}\ \emph {et~al.}(2022)\citenamefont {Lagoin},
  \citenamefont {Bhattacharya}, \citenamefont {Grass}, \citenamefont
  {Chhajlany}, \citenamefont {Salamon}, \citenamefont {Baldwin}, \citenamefont
  {Pfeiffer}, \citenamefont {Lewenstein}, \citenamefont {Holzmann},\ and\
  \citenamefont {Dubin}}]{lagoin2022checkerboard}%
  \BibitemOpen
  \bibfield  {author} {\bibinfo {author} {\bibfnamefont {C.}~\bibnamefont
  {Lagoin}}, \bibinfo {author} {\bibfnamefont {U.}~\bibnamefont
  {Bhattacharya}}, \bibinfo {author} {\bibfnamefont {T.}~\bibnamefont {Grass}},
  \bibinfo {author} {\bibfnamefont {R.~W.}\ \bibnamefont {Chhajlany}}, \bibinfo
  {author} {\bibfnamefont {T.}~\bibnamefont {Salamon}}, \bibinfo {author}
  {\bibfnamefont {K.}~\bibnamefont {Baldwin}}, \bibinfo {author} {\bibfnamefont
  {L.}~\bibnamefont {Pfeiffer}}, \bibinfo {author} {\bibfnamefont
  {M.}~\bibnamefont {Lewenstein}}, \bibinfo {author} {\bibfnamefont
  {M.}~\bibnamefont {Holzmann}},\ and\ \bibinfo {author} {\bibfnamefont
  {F.}~\bibnamefont {Dubin}},\ }\href
  {https://doi.org/10.1038/s41586-022-05123-z} {\bibfield  {journal} {\bibinfo
  {journal} {Nature}\ }\textbf {\bibinfo {volume} {609}},\ \bibinfo {pages}
  {485} (\bibinfo {year} {2022})}\BibitemShut {NoStop}%
\bibitem [{Sup()}]{Suppl}%
  \BibitemOpen
  \href@noop {} {}\bibinfo {note} {See Supplemental Material for
  additional information on perturbation theory, two-body bound state
  formation,numerical methods, calculation of the sound velocity in a lattice,
  calculation of the dynamic structure factor, and experimental implementation
  of the model, which includes
  Refs.~\cite{PhysRevA.67.053606,Pirvu2010,Haegeman2011,Haegeman2016,Pereira2008,Barthel2009,Roth2004,Iucci2006,Tokuno2011,Ejima2012,Ejima2012b,Cloizeaux1962,Yamada1969}}\BibitemShut
  {NoStop}%
\bibitem [{\citenamefont {Deuretzbacher}\ \emph {et~al.}(2010)\citenamefont
  {Deuretzbacher}, \citenamefont {Cremon},\ and\ \citenamefont
  {Reimann}}]{deuretzbacher2010ground}%
  \BibitemOpen
  \bibfield  {author} {\bibinfo {author} {\bibfnamefont {F.}~\bibnamefont
  {Deuretzbacher}}, \bibinfo {author} {\bibfnamefont {J.~C.}\ \bibnamefont
  {Cremon}},\ and\ \bibinfo {author} {\bibfnamefont {S.~M.}\ \bibnamefont
  {Reimann}},\ }\href {https://doi.org/10.1103/PhysRevA.81.063616} {\bibfield
  {journal} {\bibinfo  {journal} {Phys. Rev. A}\ }\textbf {\bibinfo {volume}
  {81}},\ \bibinfo {pages} {063616} (\bibinfo {year} {2010})}\BibitemShut
  {NoStop}%
\bibitem [{\citenamefont {Scott}\ \emph {et~al.}(1994)\citenamefont {Scott},
  \citenamefont {Eilbeck},\ and\ \citenamefont {Gilhøj}}]{SCOTT1994Sol}%
  \BibitemOpen
  \bibfield  {author} {\bibinfo {author} {\bibfnamefont {A.}~\bibnamefont
  {Scott}}, \bibinfo {author} {\bibfnamefont {J.}~\bibnamefont {Eilbeck}},\
  and\ \bibinfo {author} {\bibfnamefont {H.}~\bibnamefont {Gilhøj}},\ }\href
  {https://doi.org/https://doi.org/10.1016/0167-2789(94)90115-5} {\bibfield
  {journal} {\bibinfo  {journal} {Physica D: Nonlinear Phenomena}\ }\textbf
  {\bibinfo {volume} {78}},\ \bibinfo {pages} {194} (\bibinfo {year}
  {1994})}\BibitemShut {NoStop}%
\bibitem [{Note1()}]{Note1}%
  \BibitemOpen
  \bibinfo {note} {Note that the vanishing equilibrium density ($n_0 = 0$) in
  the gas phase can be reached only in the $N_s \rightarrow \infty $ limit.
  Instead, in a finite size system, the gas always stays at some finite
  pressure. However, zero pressure can be found in finite systems for
  self-bound states (LIQ and B-MI) with an equilibrium density
  $n_0$.}\BibitemShut {Stop}%
\bibitem [{\citenamefont {Petrosyan}\ \emph {et~al.}(2007)\citenamefont
  {Petrosyan}, \citenamefont {Schmidt}, \citenamefont {Anglin},\ and\
  \citenamefont {Fleischhauer}}]{Petrosyan2007}%
  \BibitemOpen
  \bibfield  {author} {\bibinfo {author} {\bibfnamefont {D.}~\bibnamefont
  {Petrosyan}}, \bibinfo {author} {\bibfnamefont {B.}~\bibnamefont {Schmidt}},
  \bibinfo {author} {\bibfnamefont {J.~R.}\ \bibnamefont {Anglin}},\ and\
  \bibinfo {author} {\bibfnamefont {M.}~\bibnamefont {Fleischhauer}},\ }\href
  {https://doi.org/10.1103/PhysRevA.76.033606} {\bibfield  {journal} {\bibinfo
  {journal} {Phys. Rev. A}\ }\textbf {\bibinfo {volume} {76}},\ \bibinfo
  {pages} {033606} (\bibinfo {year} {2007})}\BibitemShut {NoStop}%
\bibitem [{Note2()}]{Note2}%
  \BibitemOpen
  \bibinfo {note} {Note that this holds in the regime of applicability of the
  single-band eBH. For very strong attractive dipolar interactions, one may
  need to include higher bands into consideration.}\BibitemShut {Stop}%
\bibitem [{\citenamefont {Cazalilla}(2003)}]{PhysRevA.67.053606}%
  \BibitemOpen
  \bibfield  {author} {\bibinfo {author} {\bibfnamefont {M.~A.}\ \bibnamefont
  {Cazalilla}},\ }\href {https://doi.org/10.1103/PhysRevA.67.053606} {\bibfield
   {journal} {\bibinfo  {journal} {Phys. Rev. A}\ }\textbf {\bibinfo {volume}
  {67}},\ \bibinfo {pages} {053606} (\bibinfo {year} {2003})}\BibitemShut
  {NoStop}%
\bibitem [{\citenamefont {des Cloizeaux}\ and\ \citenamefont
  {Pearson}(1962)}]{Cloizeaux1962}%
  \BibitemOpen
  \bibfield  {author} {\bibinfo {author} {\bibfnamefont {J.}~\bibnamefont {des
  Cloizeaux}}\ and\ \bibinfo {author} {\bibfnamefont {J.~J.}\ \bibnamefont
  {Pearson}},\ }\href {https://doi.org/10.1103/PhysRev.128.2131} {\bibfield
  {journal} {\bibinfo  {journal} {Phys. Rev.}\ }\textbf {\bibinfo {volume}
  {128}},\ \bibinfo {pages} {2131} (\bibinfo {year} {1962})}\BibitemShut
  {NoStop}%
\bibitem [{\citenamefont {Yamada}(1969)}]{Yamada1969}%
  \BibitemOpen
  \bibfield  {author} {\bibinfo {author} {\bibfnamefont {T.}~\bibnamefont
  {Yamada}},\ }\href {https://doi.org/10.1143/PTP.41.880} {\bibfield  {journal}
  {\bibinfo  {journal} {Progress of Theoretical Physics}\ }\textbf {\bibinfo
  {volume} {41}},\ \bibinfo {pages} {880} (\bibinfo {year} {1969})}\BibitemShut
  {NoStop}%
\bibitem [{\citenamefont {Iucci}\ \emph {et~al.}(2006)\citenamefont {Iucci},
  \citenamefont {Cazalilla}, \citenamefont {Ho},\ and\ \citenamefont
  {Giamarchi}}]{Iucci2006}%
  \BibitemOpen
  \bibfield  {author} {\bibinfo {author} {\bibfnamefont {A.}~\bibnamefont
  {Iucci}}, \bibinfo {author} {\bibfnamefont {M.~A.}\ \bibnamefont
  {Cazalilla}}, \bibinfo {author} {\bibfnamefont {A.~F.}\ \bibnamefont {Ho}},\
  and\ \bibinfo {author} {\bibfnamefont {T.}~\bibnamefont {Giamarchi}},\ }\href
  {https://doi.org/10.1103/PhysRevA.73.041608} {\bibfield  {journal} {\bibinfo
  {journal} {Phys. Rev. A}\ }\textbf {\bibinfo {volume} {73}},\ \bibinfo
  {pages} {041608} (\bibinfo {year} {2006})}\BibitemShut {NoStop}%
\bibitem [{\citenamefont {Tokuno}\ and\ \citenamefont
  {Giamarchi}(2011)}]{Tokuno2011}%
  \BibitemOpen
  \bibfield  {author} {\bibinfo {author} {\bibfnamefont {A.}~\bibnamefont
  {Tokuno}}\ and\ \bibinfo {author} {\bibfnamefont {T.}~\bibnamefont
  {Giamarchi}},\ }\href {https://doi.org/10.1103/PhysRevLett.106.205301}
  {\bibfield  {journal} {\bibinfo  {journal} {Phys. Rev. Lett.}\ }\textbf
  {\bibinfo {volume} {106}},\ \bibinfo {pages} {205301} (\bibinfo {year}
  {2011})}\BibitemShut {NoStop}%
\bibitem [{\citenamefont {Ejima}\ \emph
  {et~al.}(2012{\natexlab{a}})\citenamefont {Ejima}, \citenamefont {Fehske},\
  and\ \citenamefont {Gebhard}}]{Ejima2012}%
  \BibitemOpen
  \bibfield  {author} {\bibinfo {author} {\bibfnamefont {S.}~\bibnamefont
  {Ejima}}, \bibinfo {author} {\bibfnamefont {H.}~\bibnamefont {Fehske}},\ and\
  \bibinfo {author} {\bibfnamefont {F.}~\bibnamefont {Gebhard}},\ }\href
  {https://doi.org/10.1088/1742-6596/391/1/012143} {\bibfield  {journal}
  {\bibinfo  {journal} {Journal of Physics: Conference Series}\ }\textbf
  {\bibinfo {volume} {391}},\ \bibinfo {pages} {012143} (\bibinfo {year}
  {2012}{\natexlab{a}})}\BibitemShut {NoStop}%
\bibitem [{\citenamefont {Ejima}\ \emph
  {et~al.}(2012{\natexlab{b}})\citenamefont {Ejima}, \citenamefont {Fehske},
  \citenamefont {Gebhard}, \citenamefont {zu~M\"unster}, \citenamefont {Knap},
  \citenamefont {Arrigoni},\ and\ \citenamefont {von~der Linden}}]{Ejima2012b}%
  \BibitemOpen
  \bibfield  {author} {\bibinfo {author} {\bibfnamefont {S.}~\bibnamefont
  {Ejima}}, \bibinfo {author} {\bibfnamefont {H.}~\bibnamefont {Fehske}},
  \bibinfo {author} {\bibfnamefont {F.}~\bibnamefont {Gebhard}}, \bibinfo
  {author} {\bibfnamefont {K.}~\bibnamefont {zu~M\"unster}}, \bibinfo {author}
  {\bibfnamefont {M.}~\bibnamefont {Knap}}, \bibinfo {author} {\bibfnamefont
  {E.}~\bibnamefont {Arrigoni}},\ and\ \bibinfo {author} {\bibfnamefont
  {W.}~\bibnamefont {von~der Linden}},\ }\href
  {https://doi.org/10.1103/PhysRevA.85.053644} {\bibfield  {journal} {\bibinfo
  {journal} {Phys. Rev. A}\ }\textbf {\bibinfo {volume} {85}},\ \bibinfo
  {pages} {053644} (\bibinfo {year} {2012}{\natexlab{b}})}\BibitemShut
  {NoStop}%
\bibitem [{\citenamefont {Landig}\ \emph {et~al.}(2015)\citenamefont {Landig},
  \citenamefont {Brennecke}, \citenamefont {Mottl}, \citenamefont {Donner},\
  and\ \citenamefont {Esslinger}}]{Landig2015}%
  \BibitemOpen
  \bibfield  {author} {\bibinfo {author} {\bibfnamefont {R.}~\bibnamefont
  {Landig}}, \bibinfo {author} {\bibfnamefont {F.}~\bibnamefont {Brennecke}},
  \bibinfo {author} {\bibfnamefont {R.}~\bibnamefont {Mottl}}, \bibinfo
  {author} {\bibfnamefont {T.}~\bibnamefont {Donner}},\ and\ \bibinfo {author}
  {\bibfnamefont {T.}~\bibnamefont {Esslinger}},\ }\href
  {https://doi.org/10.1038/ncomms8046} {\bibfield  {journal} {\bibinfo
  {journal} {Nat Commun}\ }\textbf {\bibinfo {volume} {6}},\ \bibinfo {pages}
  {7046} (\bibinfo {year} {2015})}\BibitemShut {NoStop}%
\bibitem [{\citenamefont {Cheneau}\ \emph {et~al.}(2012)\citenamefont
  {Cheneau}, \citenamefont {Barmettler}, \citenamefont {Poletti}, \citenamefont
  {Endres}, \citenamefont {Schau{\ss}}, \citenamefont {Fukuhara}, \citenamefont
  {Gross}, \citenamefont {Bloch}, \citenamefont {Kollath},\ and\ \citenamefont
  {Kuhr}}]{Bloch2012}%
  \BibitemOpen
  \bibfield  {author} {\bibinfo {author} {\bibfnamefont {M.}~\bibnamefont
  {Cheneau}}, \bibinfo {author} {\bibfnamefont {P.}~\bibnamefont {Barmettler}},
  \bibinfo {author} {\bibfnamefont {D.}~\bibnamefont {Poletti}}, \bibinfo
  {author} {\bibfnamefont {M.}~\bibnamefont {Endres}}, \bibinfo {author}
  {\bibfnamefont {P.}~\bibnamefont {Schau{\ss}}}, \bibinfo {author}
  {\bibfnamefont {T.}~\bibnamefont {Fukuhara}}, \bibinfo {author}
  {\bibfnamefont {C.}~\bibnamefont {Gross}}, \bibinfo {author} {\bibfnamefont
  {I.}~\bibnamefont {Bloch}}, \bibinfo {author} {\bibfnamefont
  {C.}~\bibnamefont {Kollath}},\ and\ \bibinfo {author} {\bibfnamefont
  {S.}~\bibnamefont {Kuhr}},\ }\href {https://doi.org/10.1038/nature10748}
  {\bibfield  {journal} {\bibinfo  {journal} {Nature}\ }\textbf {\bibinfo
  {volume} {481}},\ \bibinfo {pages} {484} (\bibinfo {year}
  {2012})}\BibitemShut {NoStop}%
\bibitem [{\citenamefont {Caux}\ and\ \citenamefont
  {Calabrese}(2006)}]{CauxCalabrese2006}%
  \BibitemOpen
  \bibfield  {author} {\bibinfo {author} {\bibfnamefont {J.-S.}\ \bibnamefont
  {Caux}}\ and\ \bibinfo {author} {\bibfnamefont {P.}~\bibnamefont
  {Calabrese}},\ }\href {https://doi.org/10.1103/PhysRevA.74.031605} {\bibfield
   {journal} {\bibinfo  {journal} {Phys. Rev. A}\ }\textbf {\bibinfo {volume}
  {74}},\ \bibinfo {pages} {031605} (\bibinfo {year} {2006})}\BibitemShut
  {NoStop}%
\bibitem [{\citenamefont {Cazalilla}(2004{\natexlab{a}})}]{Cazalilla2004_1}%
  \BibitemOpen
  \bibfield  {author} {\bibinfo {author} {\bibfnamefont {M.~A.}\ \bibnamefont
  {Cazalilla}},\ }\href {https://doi.org/10.1088/0953-4075/37/7/051} {\bibfield
   {journal} {\bibinfo  {journal} {Journal of Physics B: Atomic, Molecular and
  Optical Physics}\ }\textbf {\bibinfo {volume} {37}},\ \bibinfo {pages} {S1}
  (\bibinfo {year} {2004}{\natexlab{a}})}\BibitemShut {NoStop}%
\bibitem [{\citenamefont {Cazalilla}(2004{\natexlab{b}})}]{Cazalilla2004_2}%
  \BibitemOpen
  \bibfield  {author} {\bibinfo {author} {\bibfnamefont {M.~A.}\ \bibnamefont
  {Cazalilla}},\ }\href {https://doi.org/10.1103/PhysRevA.70.041604} {\bibfield
   {journal} {\bibinfo  {journal} {Phys. Rev. A}\ }\textbf {\bibinfo {volume}
  {70}},\ \bibinfo {pages} {041604} (\bibinfo {year}
  {2004}{\natexlab{b}})}\BibitemShut {NoStop}%
\bibitem [{\citenamefont {Anderson}\ \emph {et~al.}(2019)\citenamefont
  {Anderson}, \citenamefont {Wang}, \citenamefont {Xu}, \citenamefont {Venu},
  \citenamefont {Trotzky}, \citenamefont {Chevy},\ and\ \citenamefont
  {Thywissen}}]{Anderson2019}%
  \BibitemOpen
  \bibfield  {author} {\bibinfo {author} {\bibfnamefont {R.}~\bibnamefont
  {Anderson}}, \bibinfo {author} {\bibfnamefont {F.}~\bibnamefont {Wang}},
  \bibinfo {author} {\bibfnamefont {P.}~\bibnamefont {Xu}}, \bibinfo {author}
  {\bibfnamefont {V.}~\bibnamefont {Venu}}, \bibinfo {author} {\bibfnamefont
  {S.}~\bibnamefont {Trotzky}}, \bibinfo {author} {\bibfnamefont
  {F.}~\bibnamefont {Chevy}},\ and\ \bibinfo {author} {\bibfnamefont {J.~H.}\
  \bibnamefont {Thywissen}},\ }\href
  {https://doi.org/10.1103/PhysRevLett.122.153602} {\bibfield  {journal}
  {\bibinfo  {journal} {Phys. Rev. Lett.}\ }\textbf {\bibinfo {volume} {122}},\
  \bibinfo {pages} {153602} (\bibinfo {year} {2019})}\BibitemShut {NoStop}%
\bibitem [{\citenamefont {Krutitsky}(2016)}]{KRUTITSKY2016}%
  \BibitemOpen
  \bibfield  {author} {\bibinfo {author} {\bibfnamefont {K.~V.}\ \bibnamefont
  {Krutitsky}},\ }\href
  {https://doi.org/https://doi.org/10.1016/j.physrep.2015.10.004} {\bibfield
  {journal} {\bibinfo  {journal} {Phys. Rep.}\ }\textbf {\bibinfo {volume}
  {607}},\ \bibinfo {pages} {1} (\bibinfo {year} {2016})}\BibitemShut {NoStop}%
\bibitem [{\citenamefont {Pirvu}\ \emph {et~al.}(2010)\citenamefont {Pirvu},
  \citenamefont {Murg}, \citenamefont {Cirac},\ and\ \citenamefont
  {Verstraete}}]{Pirvu2010}%
  \BibitemOpen
  \bibfield  {author} {\bibinfo {author} {\bibfnamefont {B.}~\bibnamefont
  {Pirvu}}, \bibinfo {author} {\bibfnamefont {V.}~\bibnamefont {Murg}},
  \bibinfo {author} {\bibfnamefont {J.~I.}\ \bibnamefont {Cirac}},\ and\
  \bibinfo {author} {\bibfnamefont {F.}~\bibnamefont {Verstraete}},\ }\href
  {https://doi.org/10.1088/1367-2630/12/2/025012} {\bibfield  {journal}
  {\bibinfo  {journal} {New Journal of Physics}\ }\textbf {\bibinfo {volume}
  {12}},\ \bibinfo {pages} {025012} (\bibinfo {year} {2010})}\BibitemShut
  {NoStop}%
\bibitem [{\citenamefont {Haegeman}\ \emph {et~al.}(2011)\citenamefont
  {Haegeman}, \citenamefont {Cirac}, \citenamefont {Osborne}, \citenamefont
  {Pi\ifmmode~\check{z}\else \v{z}\fi{}orn}, \citenamefont {Verschelde},\ and\
  \citenamefont {Verstraete}}]{Haegeman2011}%
  \BibitemOpen
  \bibfield  {author} {\bibinfo {author} {\bibfnamefont {J.}~\bibnamefont
  {Haegeman}}, \bibinfo {author} {\bibfnamefont {J.~I.}\ \bibnamefont {Cirac}},
  \bibinfo {author} {\bibfnamefont {T.~J.}\ \bibnamefont {Osborne}}, \bibinfo
  {author} {\bibfnamefont {I.}~\bibnamefont {Pi\ifmmode~\check{z}\else
  \v{z}\fi{}orn}}, \bibinfo {author} {\bibfnamefont {H.}~\bibnamefont
  {Verschelde}},\ and\ \bibinfo {author} {\bibfnamefont {F.}~\bibnamefont
  {Verstraete}},\ }\href {https://doi.org/10.1103/PhysRevLett.107.070601}
  {\bibfield  {journal} {\bibinfo  {journal} {Phys. Rev. Lett.}\ }\textbf
  {\bibinfo {volume} {107}},\ \bibinfo {pages} {070601} (\bibinfo {year}
  {2011})}\BibitemShut {NoStop}%
\bibitem [{\citenamefont {Haegeman}\ \emph {et~al.}(2016)\citenamefont
  {Haegeman}, \citenamefont {Lubich}, \citenamefont {Oseledets}, \citenamefont
  {Vandereycken},\ and\ \citenamefont {Verstraete}}]{Haegeman2016}%
  \BibitemOpen
  \bibfield  {author} {\bibinfo {author} {\bibfnamefont {J.}~\bibnamefont
  {Haegeman}}, \bibinfo {author} {\bibfnamefont {C.}~\bibnamefont {Lubich}},
  \bibinfo {author} {\bibfnamefont {I.}~\bibnamefont {Oseledets}}, \bibinfo
  {author} {\bibfnamefont {B.}~\bibnamefont {Vandereycken}},\ and\ \bibinfo
  {author} {\bibfnamefont {F.}~\bibnamefont {Verstraete}},\ }\href
  {https://doi.org/10.1103/PhysRevB.94.165116} {\bibfield  {journal} {\bibinfo
  {journal} {Phys. Rev. B}\ }\textbf {\bibinfo {volume} {94}},\ \bibinfo
  {pages} {165116} (\bibinfo {year} {2016})}\BibitemShut {NoStop}%
\bibitem [{\citenamefont {Pereira}\ \emph {et~al.}(2008)\citenamefont
  {Pereira}, \citenamefont {White},\ and\ \citenamefont
  {Affleck}}]{Pereira2008}%
  \BibitemOpen
  \bibfield  {author} {\bibinfo {author} {\bibfnamefont {R.~G.}\ \bibnamefont
  {Pereira}}, \bibinfo {author} {\bibfnamefont {S.~R.}\ \bibnamefont {White}},\
  and\ \bibinfo {author} {\bibfnamefont {I.}~\bibnamefont {Affleck}},\ }\href
  {https://doi.org/10.1103/PhysRevLett.100.027206} {\bibfield  {journal}
  {\bibinfo  {journal} {Phys. Rev. Lett.}\ }\textbf {\bibinfo {volume} {100}},\
  \bibinfo {pages} {027206} (\bibinfo {year} {2008})}\BibitemShut {NoStop}%
\bibitem [{\citenamefont {Barthel}\ \emph {et~al.}(2009)\citenamefont
  {Barthel}, \citenamefont {Schollw\"ock},\ and\ \citenamefont
  {White}}]{Barthel2009}%
  \BibitemOpen
  \bibfield  {author} {\bibinfo {author} {\bibfnamefont {T.}~\bibnamefont
  {Barthel}}, \bibinfo {author} {\bibfnamefont {U.}~\bibnamefont
  {Schollw\"ock}},\ and\ \bibinfo {author} {\bibfnamefont {S.~R.}\ \bibnamefont
  {White}},\ }\href {https://doi.org/10.1103/PhysRevB.79.245101} {\bibfield
  {journal} {\bibinfo  {journal} {Phys. Rev. B}\ }\textbf {\bibinfo {volume}
  {79}},\ \bibinfo {pages} {245101} (\bibinfo {year} {2009})}\BibitemShut
  {NoStop}%
\bibitem [{\citenamefont {Roth}\ and\ \citenamefont
  {Burnett}(2004)}]{Roth2004}%
  \BibitemOpen
  \bibfield  {author} {\bibinfo {author} {\bibfnamefont {R.}~\bibnamefont
  {Roth}}\ and\ \bibinfo {author} {\bibfnamefont {K.}~\bibnamefont {Burnett}},\
  }\href {https://doi.org/10.1088/0953-4075/37/19/009} {\bibfield  {journal}
  {\bibinfo  {journal} {Journal of Physics B: Atomic, Molecular and Optical
  Physics}\ }\textbf {\bibinfo {volume} {37}},\ \bibinfo {pages} {3893}
  (\bibinfo {year} {2004})}\BibitemShut {NoStop}%
\end{thebibliography}%

\pagebreak
\onecolumngrid
\vspace{\columnsep}
\newpage
\begin{center}
\textbf{\large Supplementary Material for 'Superexchange liquefaction of strongly correlated lattice dipolar bosons'}
\end{center}
\vspace{2cm}
\twocolumngrid

\setcounter{equation}{0}
\setcounter{figure}{0}
\makeatletter
\renewcommand{\theequation}{S\arabic{equation}}
\addtolength{\textfloatsep}{5mm}
\section{Lattice fermions with a long-range interaction}
One can obtain a correct description of relevant features of the system by perturbatively adding an attractive long-range interaction to the non-interacting lattice Fermi gas. By considering the Fermi gas state we can estimate the energy as,
\begin{align}
    E &= E_K+E_V \\
    &= -2JN_s \sin(n\pi)/\pi + V \sum_{i<j=1}^{N_s} \frac{\langle \hat{n}_i \hat{n}_j \rangle}{|i-j|^{3}}.\nonumber
\end{align}
The first term represents the kinetic energy of the lattice Fermi gas that we obtain in the thermodynamic limit $N,N_s\rightarrow \infty$ for a fixed particle density $n=N/N_s$. The second term is the dipolar energy, which can be computed in a perturbative manner by employing Wick's theorem $\langle \hat{n}_i \hat{n}_j \rangle/n^2 = 1-\sin^2(k_F|i-j|)/(k_F|i-j|)^2$, where we introduce the Fermi momentum $k_F=\pi n$, and we employ $\langle\hat{c}^{\dagger}_i\hat{c}_j\rangle = \sin(k_F |i-j|)/(\pi|i-j|)$. After some algebraic manipulations, we obtain the total energy per particle,
\begin{align}
    E/N &= -2J \sin(n\pi)/(n\pi) + V \xi(3) n \label{Eq:EoS1}\\ 
    &- \frac{V \xi(5)}{2n\pi^2} \nonumber \\
    &+\frac{V}{4n\pi^2}\left[\textrm{Li}_{5}\left(e^{2 i \pi n}\right)+\textrm{Li}_{5}\left(e^{-2 i \pi n}\right) \right].\nonumber
\end{align}
The liquid phase becomes energetically favorable when the energy per particle is smaller as compared to its value in the gas phase at zero density, that is when $E/N < -2J$. In principle, the attractive long-range interaction lowers the energy per particle and the system will liquefy for a critical value of $V/J$. Since we deal with hard-core particles, the system cannot have more than one particle per site. This imposes limitations on the equilibrium density, which should lie in the $0<n_0\leq 1$ range. Therefore, there is a critical point of the long-range interaction strength for which the equilibrium density of the liquid corresponds to the occupation of one particle per site. In that situation, self-bound Mott insulators appear in finite-size systems while bulk systems exhibit insulator properties and are completely incompressible. The perturbative equation of state Eq.~\eqref{Eq:EoS1} allows for estimating a critical value for the long-range strength at which the equilibrium density touches one,
\begin{align}
    V_{\textrm{MI}}/J = 2/\xi(3).
    \label{Eq:V_MI}
\end{align}
To see the appearance of a liquid phase, we need to obtain $E/N<-2J$ before reaching the self-bound Mott-insulator phase $V>V_{\textrm{MI}}$. By evaluating the energy per particle at the transition point we obtain $E/N|_{n=1,V=V_{\textrm{MI}}}=-2J$. Therefore, we see that there is no intermediate regime where a liquid phase can be found with $E/N<-2J$ and $0<n_0<1$. The perturbative equation of state Eq.~\eqref{Eq:EoS1} predicts a direct transition between a gas and a self-bound Mott-insulator at a strength given by Eq.~\eqref{Eq:V_MI}.
\begin{figure}[t!]
\centering
\includegraphics[width=1\columnwidth]{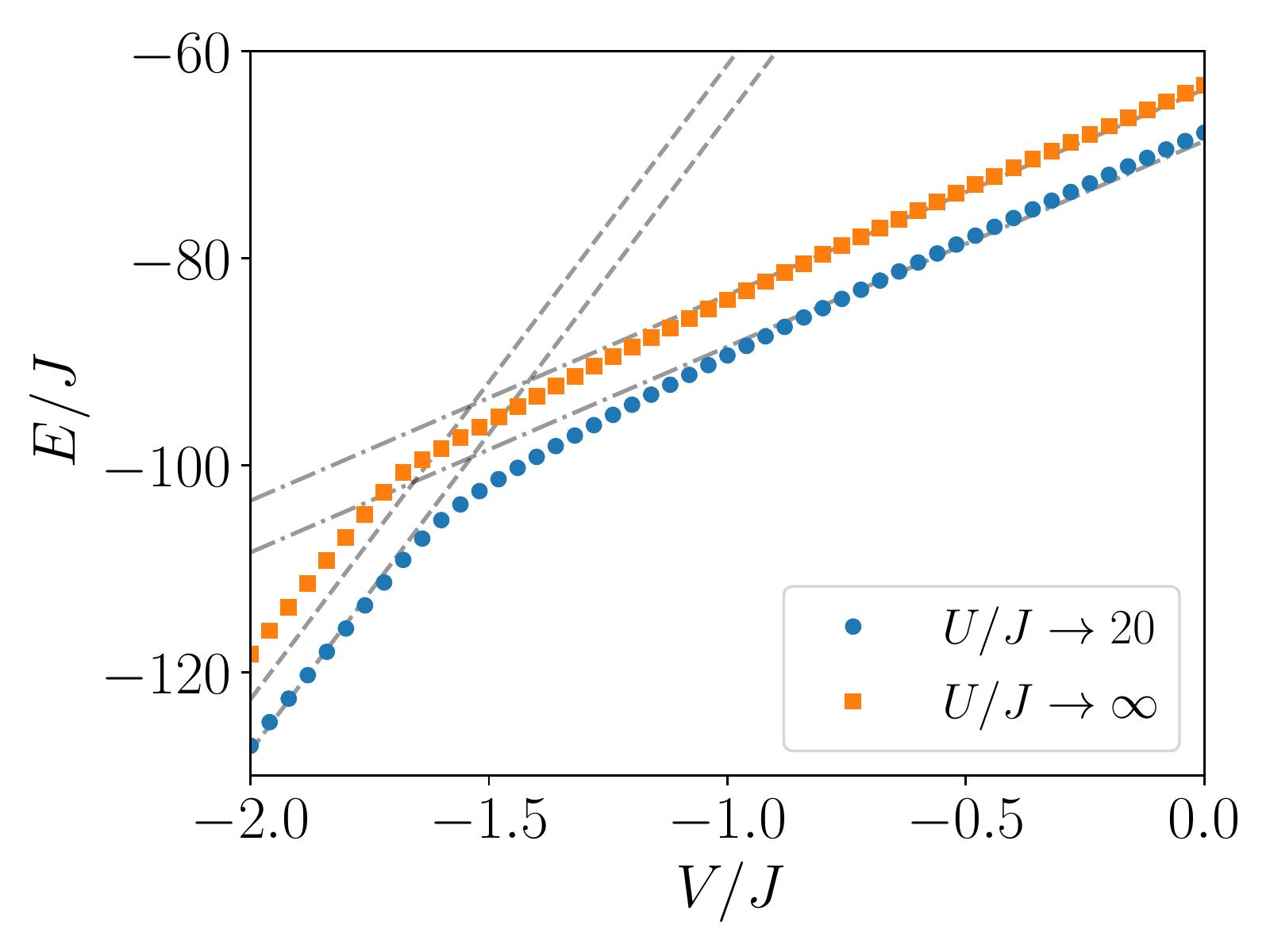}
\caption{Energy $E$ of dipolar bosons loaded in a one-dimensional optical lattice as a function of the dipolar strength $V$ for the case of hard-core bosons $U/J\rightarrow \infty$ and a strong but finite on-site interaction $U/J=20$. The simulations are performed with $N=50$ particles in a lattice of $N_s=100$ sites. Perturbative energy calculations for a fixed density $n=1$ (dashed line) and $n=0.5$ (dotted-dashed line).}
\label{Fig:EnergyPert}
\end{figure}
\section{Quasi-Tonks-Girardeau regime with on-site and long-range interactions}
After examining the upshots of adding long-range interaction to an ideal Fermi system, we also include the effect of having realistic bosons (not hard-core ones) with strong on-site interaction $U$. This effect can be taken into account at second order in degenerate perturbation theory in $U/J$~\cite{PhysRevA.67.053606}, leading to the effective Hamiltonian,
\begin{align}
\hat{H} &= -J\sum_{i=1}^{N_s} \left(\hat{c}_i^{\dagger} \hat{c}_{i+1} + \text{h.c.} \right) +\frac{2J^2}{U}\sum_{i=1}^{N_s}  \left(\hat{c}_{i-1}^{\dagger} \hat{n}_i \hat{c}_{i+1} + \text{h.c.} \right)  \nonumber \\  
&- \frac{4J^2}{U} \sum_{i=1}^{N_s} \hat{n}_i \hat{n}_{i+1} + V\sum_{i<j=1}^{N_s} \frac{\hat{n}_i \hat{n}_j}{|i-j|^{3}},
\label{Eq:BHHC}
\end{align}
where $\hat{c}^{\dagger}_i$ and $\hat{c}_i$ represent creation and annihilation fermionic operators at site $i$. The energy of the system is given by,
\begin{equation}
    E = E_K+E_V+E_U,
    \label{Eq:Pert_E}
\end{equation}
where $E_K$ and $E_V$ are given by Eq.~\eqref{Eq:EoS1} and the on-site energy reads,
\begin{align}
    E_U/N = -\frac{4J^2}{U} n \left(1- \frac{\sin(2\pi n)}{2\pi n}  \right).
\end{align}
We observe that the reduction of infinite on-site repulsion to a finite one has an attractive effect and $E_U<0$. Therefore, the quasi-TG fluid is less repulsive than the Fermi fluid for the same long-range coupling. Due to the different density dependence of the energy contributions, the on-site interaction can liquefy the system. Thus, by reducing the on-site interaction, we expect to observe a gas-to-liquid transition.  In the dilute regime $n\ll 1$, the perturbative terms remain small compared to the rest of Eq.~\eqref{Eq:Pert_E}. On the other hand, for larger densities they play an important role. In particular, one can observe the non-monotonous behavior of the energy as a function of the density. This will have strong implications on many thermodynamic observables of the quasi-TG liquid such as its compressibility. 

To test predictions of the developed perturbative theory, we perform DMRG simulations of Hamiltonian by fixing a number of particles in the system and varying the long-range strength $V$ with a large on-site interaction $U$, see Fig.~\ref{Fig:EnergyPert}. When $V<V_{\textrm{MI}}$, see Eq.~\eqref{Eq:V_MI}, the homogeneous solution is stable and the perturbative calculations with a homogeneous density given by $n=N/N_s$ correctly predict the behavior of the energy of the system as a function of the long-range strength. When it exceeds the critical value $V>V_c$, the homogeneous solution becomes unstable with respect to the formation of a self-bound MI which is a droplet with a saturated density $n=1$. By employing the same perturbative equations for a fixed density $n=1$, we can also predict the dependence of the self-bound MI energy on the dipolar interaction strength. The abrupt change of the energy as the long-range strength reaches the critical value $V_c$ signals the presence of a first-order phase transition between the homogeneous liquid and an inhomogeneous, completely incompressible self-bound insulator.

\section{The two-body problem}
\begin{figure}[t!]
\centering
\includegraphics[width=1\columnwidth]{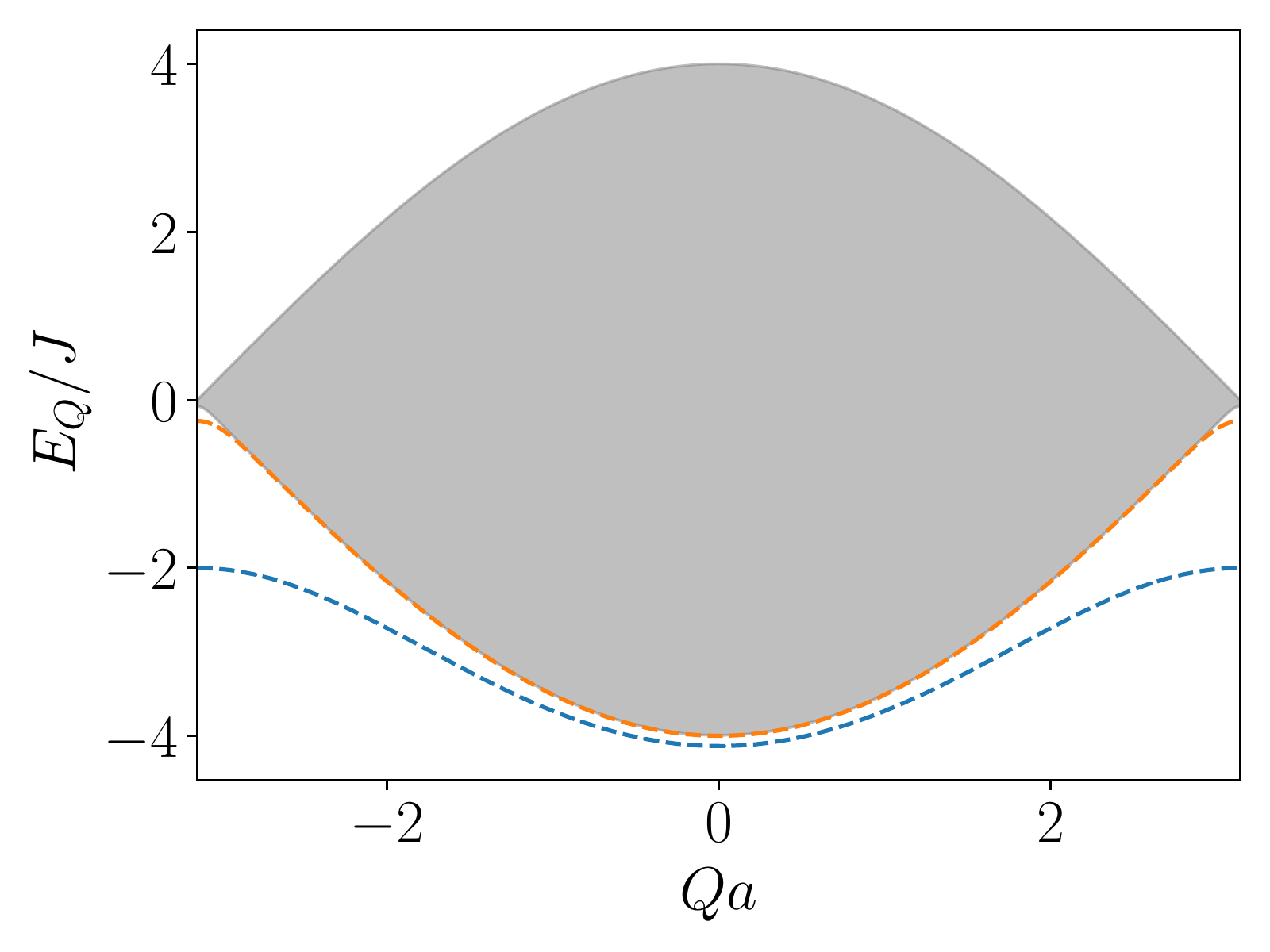}
\caption{Energy of two dipolar bosons in a 1D optical lattice as a function of the total quasi-momentum $Qa$ of the pair. The on-site interaction is $U/J=20$ and the dipolar strength $V/J=-10$. The filled gray region represents the scattering continuum and dashed lines depict bound state solutions. 
}
\label{Fig:SpectrumTwo}
\end{figure}

\begin{figure}[t!]
\centering
\includegraphics[width=1\columnwidth]{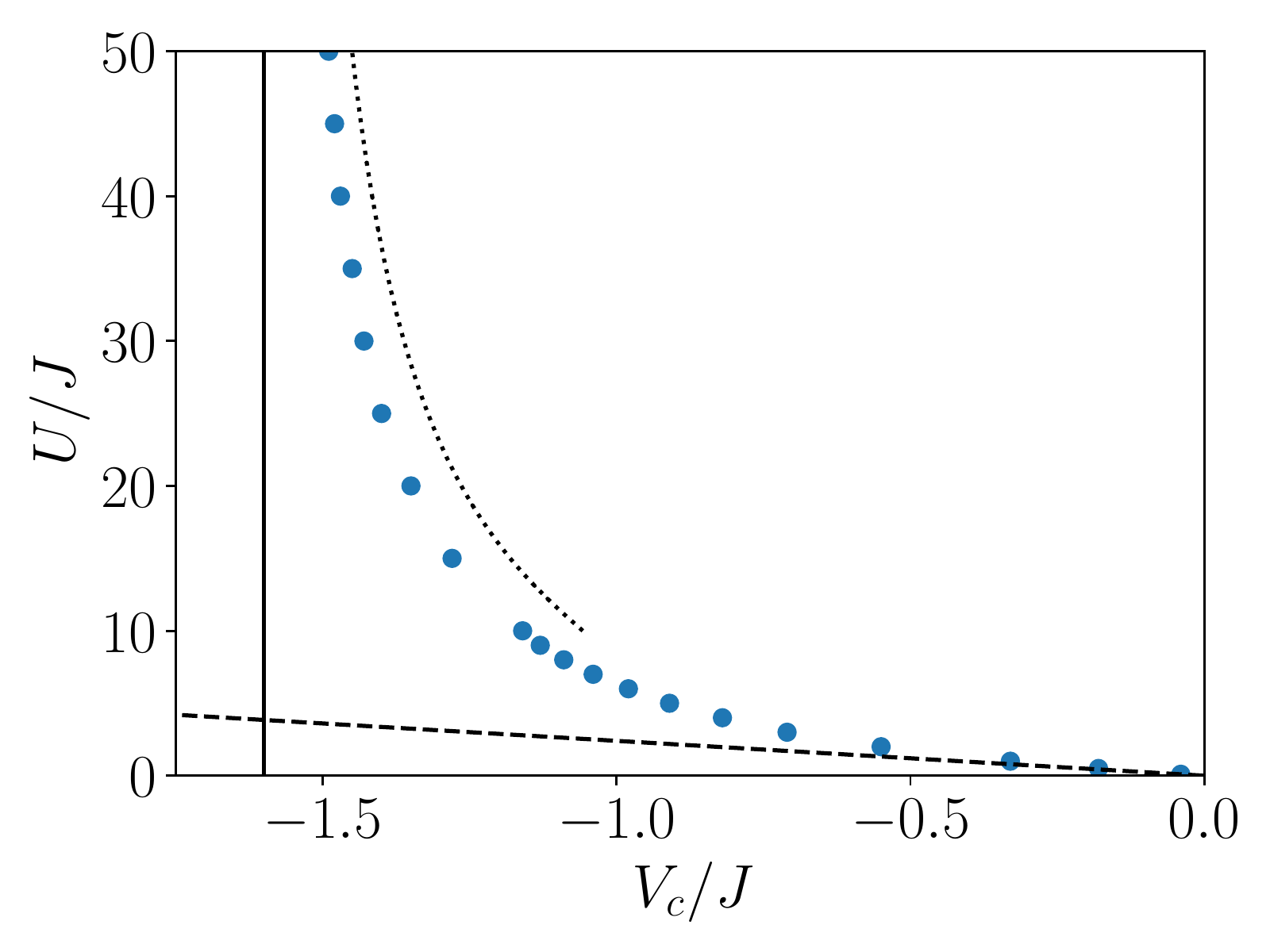}
\caption{Critical value of the dipolar strength $V_c/J$ for bound state formation for different values of the on-site interaction $U/J$. The dashed line represents the critical dipolar strength obtained in the fermionized limit $U/J\rightarrow \infty$.
}
\label{Fig:Threshold}
\end{figure}

A problem of two dipolar bosons in a one-dimensional lattice can be solved by separating the center of mass and relative motion and using the following set of states,
\begin{equation}
|\psi\rangle = \frac{1}{\sqrt{N_s}} \sum_{i,j} e^{iQR} \psi_Q(z)|i,j\rangle,
\label{Eq:TwoPart}
\end{equation}
where we introduce the total quasi-momentum $Q$ of a pair, relative $z=i-j$ and total $R=(i+j)/2$ coordinates, a wavefunction of the relative motion $\psi_Q(z)$ and the two-particle state $|i,j\rangle = \hat{b}_i^{\dagger} \hat{b}_j^{\dagger}|0\rangle$. After inserting Eq.~\eqref{Eq:TwoPart} into Hamiltonian, we obtain the equation of motion for the relative wavefunction,
\begin{align}
    E_Q \psi_Q(z) &= \sum_{e=\pm 1} -2J\cos(Q/2) \psi_Q(z+e) \label{Eq:EoM_rel}\\ 
    &+ U \delta(z)\psi_Q(z) +2V \sum_{n=1}^{N_s} \frac{\delta(|z|-n)}{n^{3}}\psi_Q(z). \nonumber
\end{align}
Above, we introduce the energy of the pair $E_Q$ that depends parametrically on the quasi-momentum $Q$. By numerically solving Eq.~\eqref{Eq:EoM_rel}, we get a spectrum of the system as a function of the pair quasi-momentum presented in Fig.~\ref{Fig:SpectrumTwo}. We notice that for large values of the on-site interaction $U\gg J$, there is a critical negative value of the dipolar strength $V_c/J$, for which a bound state appears in the spectrum. This bound state appears below the two-particle scattering continuum and its energy has a minimum at $Qa=0$. It is characterized by a negative binding energy $E_B/J$ defined as a difference between the energy of the state and the minimum of the scattering band. For larger negative values of the dipolar strength, more bound states can be found. For the deepest bound state in the system, we compute a typical relative distance between the two particles $z^*$ as a function of the dipolar strength. We observe that after crossing the critical dipolar strength $V_c/J$, the two particles already localize in adjacent sites. This indicates the local nature of the bound state even though the presence of a long-range dipolar interaction. We conclude that the typical distance between two particles is set by the lattice spacing $z^*\sim a$. We also explore the dependence of the critical dipolar strength $V_c/J$ for bound state formation on the on-site interaction $U/J$, see Fig.~\ref{Fig:Threshold}. The critical value $V_c/J$ decreases rapidly for small values of the on-site interaction $U/J$. For larger on-site repulsion, it slowly tends to the critical value $V_c/J\sim -1.61$ in the fermionization limit  ($U/J\rightarrow\infty$). This is also the critical dipolar strength obtained for the fermionic case, where the relative wavefunction $\psi_Q(z)$ is completely antisymmetric.
\section{Numerical method}

To support our analytical theory, we employ unbiased tensor network simulations. For obtaining the ground state we employ the density-matrix renormalization group (DMRG) algorithm. For performing real-time evolution we make use of the time-dependent variational principle (TDVP), see Sec.~\ref{Sec.Dyn}. 

We employ the standard two-site update DMRG and infinite-DMRG algorithms in order to find the ground state of Hamiltonian~(1). To calculate various properties of different phases exhibited by the considered system, we performed simulations with $N_s=100-800$ lattice sites and $N=50-100$ particles. To obtain the density dependence, we imposed the periodic boundary conditions on Hamiltonian~(1) and employed DMRG. We find an excellent agreement between these results and the ones obtained with the iDMRG method where the homogeneous solution is mechanically stable. To obtain the quantum droplets in Fig.~3, we used open boundary conditions ensuring that the system size does not affect the profiles. In our code, to go beyond the nearest-neighbor interaction and couple two arbitrary distant dipolar bosons, we fit the power-law decaying dipolar interaction with a sum of ten exponential functions~\cite{Pirvu2010}, when having open boundary conditions. Thus, we are not cutting the range of the interaction in the numerical implementation of our model. For systems with periodic boundary conditions we define the distance between two particles as the arc of the respective ring.
To test the MPS representation of the ground state, we have checked the convergence in terms of the bond dimension $\chi$ up to $\chi=800$. The quantities presented in the main text, such as the energy or particle variance, show a fast convergence with the bond dimension. At the same time, we have implemented a cutoff in the maximum number of bosons per site $N_m$. Our perturbative theory hints that doublons play a crucial role in the formation of the liquid state. We thus also checked the convergence for the varying cutoff $N_m$. For the case $U=20$ and $V=-1.5$, which shows the maximum particle variance (see the main text), we observe an excellent convergence already obtained for $N_m=3$. The energy and entanglement entropy difference with respect to $N_m=4$ are $\Delta E = 10^{-7}$ and $\Delta S = 10^{-7}$. Therefore, all the results presented in the main text are performed at $N_m=3$. The excellent agreement between our perturbative theory and the DMRG results shows that doublons created through superexchange processes are the main mechanism for liquefaction. 

\section{Gas to liquid transition}
\begin{figure}[t!]
\centering
\includegraphics[width=1\columnwidth]{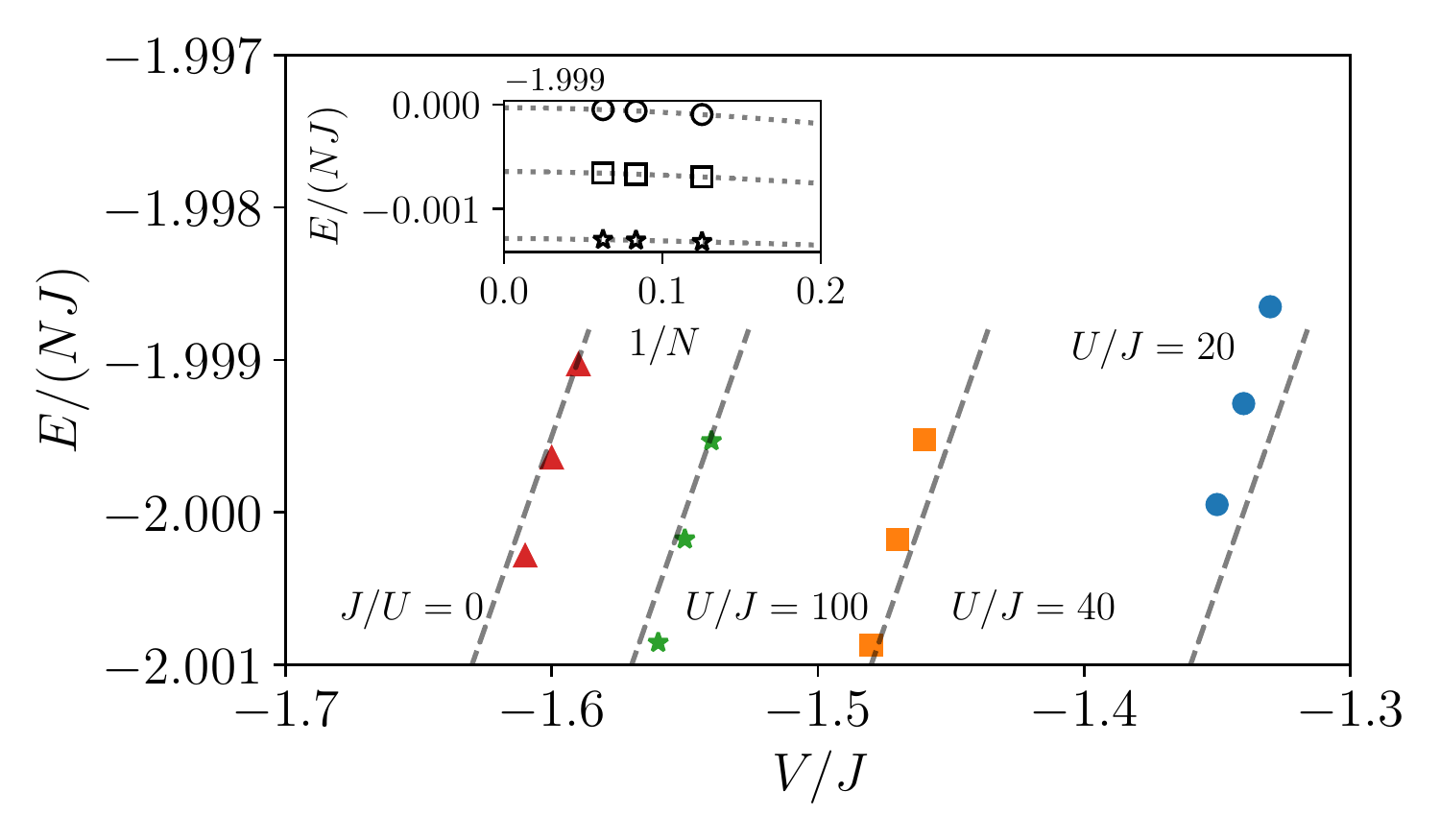}
\caption{Main panel: Energy per particle in the thermodynamic limit for a system density $n=1/10$ as a function of the dipolar strength $V/J$ for different values of on-site interaction $U/J$. The gas-to-liquid phase transition occurs when $E/(NJ)=-2$. Dashed lines indicate a two-body decoupling, see the main text. Inset panel: Energy per particle dependence with the number of particles at fixed density for the case of hard-core bosons.
}
\label{Fig:GastoLiquid}
\end{figure}
To detect the gas-to-liquid phase transition, we study the stability of the homogeneous gas as a function of the interactions for a very small density $n=0.1$. To fix the density of the system, we impose the periodic boundary conditions on the Hamiltonian and then employ DMRG. For increasing dipolar interaction, the many-body gas state becomes self-bound $E/N<-2J$ and a transition to a liquid phase occurs, see Fig.~\ref{Fig:GastoLiquid}. To ensure the many-body transition, we study the dependence of the energy per particle with the number of particles for fixed density $n=1/10$. This allows us to extrapolate the observables to the thermodynamic limit. 

The energy per particle depends almost linearly on the strength of dipolar interactions. This points to a two-body decoupling. All interaction effects can be absorbed in a two-body coupling strength $g_0 = (V-V_0)$, $V_0$ being the point at which a two-body bound state appears. The energy per particle can be thus estimated as $E/N = \frac{1}{2}g_0n$. Remarkably, this gives a good estimation of the energy per particle close to the liquid-to-gas transition, see Fig.~\ref{Fig:GastoLiquid}.

\section{Particle variance}
The local particle variance can be directly related with the local pair-correlation function $g_2=\langle \hat{b}^{\dagger}\hat{b}^{\dagger}\hat{b}\hat{b}\rangle/n^2$,
\begin{equation}
    \Delta n \equiv \sqrt{\langle \hat{n}^2\rangle - n ^2} = \sqrt{g_2 n^2+n-n^2}. 
\end{equation}
By employing the Hellmann-Feynman theorem, the local pair-correlation function can be written as,
\begin{equation}
    g_2 = \frac{2}{n^2} \frac{\partial}{\partial U} \left(\frac{E_U}{N_s} \right)=\frac{8J^2 }{U^2 n^2}\left(1-\frac{\sin(2 \pi n)}{2 \pi n} \right),
\end{equation}
which leads to particle variance,
\begin{equation}
    \Delta n \equiv  \sqrt{n-n^2 + \frac{8J^2 }{U^2}\left(1-\frac{\sin(2\pi n)}{2 \pi n} \right)}.
\end{equation}
The particle variance can be used to detect different phases present in the system. The gas-to-liquid phase transition is characterized by a discontinuity in the first derivative of the local particle variance at equilibrium $\Delta n_0 \sim \sqrt{n_0}$. On the other hand, when crossing the transition from liquid to self-bound Mott-insulator, the particle variance is an analytic differentiable function $\Delta n_0\sim 2\sqrt{2} J/\sqrt{U} - (n_0-1)\frac{8J^2+U}{4\sqrt{2}J \sqrt{U}}$ for any value of finite on-site interaction $U/J\gg 1$. However, it shows a kink at the transition point owing to the presence of a discontinuity of the first derivative of the equilibrium density. In the Tonks-Girardeau limit $U/J\rightarrow \infty$, the particle variance also exhibits a discontinuity in its first derivative $\Delta n_0 \sim \sqrt{1-n_0}$. 

\section{Density and phase response}
Lattices naturally break Galilean invariance since the particle current is not a conserved quantity in any two-particle collision even though the total quasi-momentum is conserved. This has dramatic consequences on many thermodynamic and transport observables. It also implies that the low-energy Luttinger theory cannot be described by a single parameter but rather needs two distinct phenomenologic parameters. These two parameters can be introduced by measuring the response of the system to a change in particle number and to a phase shift,
\begin{equation}
    v_N = \frac{N_sa}{\pi} \frac{\partial^2 E}{ \partial N^2}\Big|_{N=N_0}, \quad v_J = \pi N_s a \frac{\partial^2 E}{ \partial \phi^2}\Big|_{\phi=0}.
\end{equation}
Notice that for a Galilean invariant system, one parameter is immediately reduced to the Fermi velocity $v_J=v_F$, which is fully fixed by the linear density $v_F=\pi n$. In that case, the low-energy properties are entirely characterized by the adiabatic compressibility. Instead, for a lattice system, one has to compute both parameters in a microscopic calculation. The low-energy properties of the fluid can be expressed in terms of these parameters,
\begin{equation}
    K=\sqrt{\frac{v_J}{v_N}}, \quad v_s = \sqrt{v_J v_N},
    \label{Eq:LuttPar}
\end{equation}
$v_s$ being the speed of sound of the low-energy phonons and $K$ the Luttinger parameter determining the long-range correlations in the fluid. The velocity $v_N$ is inversely proportional to the compressibility of the system showing that the more incompressible the fluid, the larger the velocity $v_N$. On the other hand, the speed of sound $v_s$ is also affected by the response to phase fluctuations and both $v_J$ and $v_N$ have an effect on it. The compressibility can already be computed from Eq.~\eqref{Eq:Pert_E} leading to the density response,
\begin{align}
    v_N &= 2J\sin(\pi n)+\frac{8 J^2}{\pi U}\left(\cos(2\pi n)- n \sin(2\pi n) -1  \right) \nonumber \\
    &+\frac{V}{\pi}\left[2\xi(3)+\textrm{Li}_{3}\left(e^{-2 i \pi n}\right)+\textrm{Li}_{3}\left(e^{2 i \pi n}\right) \right].
\end{align}
To compute the response to phase fluctuations, we also performed a perturbative calculation imposing twisted boundary conditions $\hat{c}_{i+N_s}=e^{i\phi}\hat{c}_i$. In the thermodynamic limit, this leads to a phase shift in the correlation function $\langle\hat{c}^{\dagger}_i\hat{c}_j\rangle = e^{i\phi |i-j|/N_s} \sin(k_F |i-j|)/(\pi|i-j|)$. By employing Wick's theorem, we can again obtain perturbatively the dependence of the energy on the phase $\phi$. After some manipulations, the phase response is given by,
\begin{align}
    v_J = 2J \sin(\pi n) + \frac{8J^2}{\pi U}\left[ 2\sin^2(\pi n) - \pi n \sin(2\pi n ) \right]
    \label{Eq:vJ}
\end{align}
In our perturbative calculation, we observe that the dipolar interaction does not lead to any modification of the phase response in the system. This is already expected since it is a density-density interaction, and it only affects the compressibility of the system. 

Combining our calculations, we are able to calculate the speed of sound in the lattice TG with a non-local interaction,
\begin{align}
    v_s &= 2J\sin(\pi n)-\frac{16 J^2}{U}n\sin(\pi n) \cos(\pi n) \nonumber \\
    &+\frac{V}{2\pi}\left[2\xi(3)+\textrm{Li}_{3}\left(e^{-2 i \pi n}\right)+\textrm{Li}_{3}\left(e^{2 i \pi n}\right) \right].
\end{align}

\section{Feynman relation in a lattice}
In absence of an optical lattice, the Feynman relation provides an upper bound to the lower branch of the excitation spectrum $\omega_-(k)$, in terms of the static structure factor $S(k)$ according to $\omega_-(k)\leq \hbar^2k^2/[2mS(k)]$. 
The bound becomes exact when excitations are exhausted by a single mode. It happens, for example, at low momenta in compressible phases where linear phonons are strongly populated, that is in the regime in which the Luttinger Liquid description applies. The lattice analog of the Feynman relation can be found by employing the non-Galilean invariant Luttinger theory. The long-distance decay of density-density correlation is then dictated by the Luttinger parameters Eq.~\eqref{Eq:LuttPar},
\begin{equation}
    \langle \delta n_i \delta n_j \rangle \sim  \frac{K}{2\pi^2 |i-j|^2},
    \quad |i-j|\to\infty
    \label{Eq:DensDec}
\end{equation}
where $\delta n_i= n_i - n$. The pair correlation function is defined as an expectation value of the density-density distribution, 
\begin{equation}
g = 1+\frac{1}{n^2} \langle \delta n_i \delta n_j \rangle.
\end{equation}
The static structure factor quantifies two-body correlations in momentum space and can be obtained from the pair correlation function by using Fourier transform,
\begin{equation}
S(k) = 1 + n\sum_{j} e^{ik j} (g-1).
\end{equation}
The long-distance inverse-square decay of the density-density correlations~\eqref{Eq:DensDec} results in a linear low-momentum dependence of the static structure factor,
\begin{equation}
S(k) = \frac{K}{2n\pi}k,\quad |k|\to 0.
\label{Eq:Sk:phonons}
\end{equation}
By using the thermodynamic relation for the compressibility $\kappa$ and using Eq.~\eqref{Eq:LuttPar}, we obtain the following relations
\begin{equation}
    \kappa_s^{-1} = n^2\frac{\partial \mu}{\partial n}= \pi v_Nn^2=\frac{ \pi v_s n^2}{K},
\end{equation}
between the inverse compressibility $\kappa_s^{-1}$, the speed of sound $v_s$ and the Luttinger parameter $K$. By inserting these relations into the low-momentum expansion of the static structure factor, Eq.~(\ref{Eq:Sk:phonons}),  we obtain,
\begin{equation}
S(k) = \frac{\kappa_s v_s n}{2}k.
\label{Eq:Sk:Feynman}
\end{equation}
The above is the lattice analog of the Feynman relation in the continuum. By knowing the low-momentum behavior of the static structure factor and the compressibility of the system, we can calculate the speed of sound and hence the phononic part of the excitation spectrum, $\omega(k) = v_s k $.


To recover the continuum limit, one just has to set $v_J=v_F=\pi n$, which sets the inverse compressibility $\kappa_s^{-1}= v_s^2 n$ and thus obtaining 
\begin{equation}
S(k) = \frac{k}{2v_s},
\end{equation}
which corresponds to the Feynman relation in the continuum with $\hbar=m=1$. In the continuum, one thus has a direct relationship between compressibility and the speed of sound. Namely, when a system becomes less compressible, its speed of 
sound increases.

\section{Dynamic structure factor}
\label{Sec.Dyn}
\begin{figure}[t!]
\centering
\includegraphics[width=1\columnwidth]{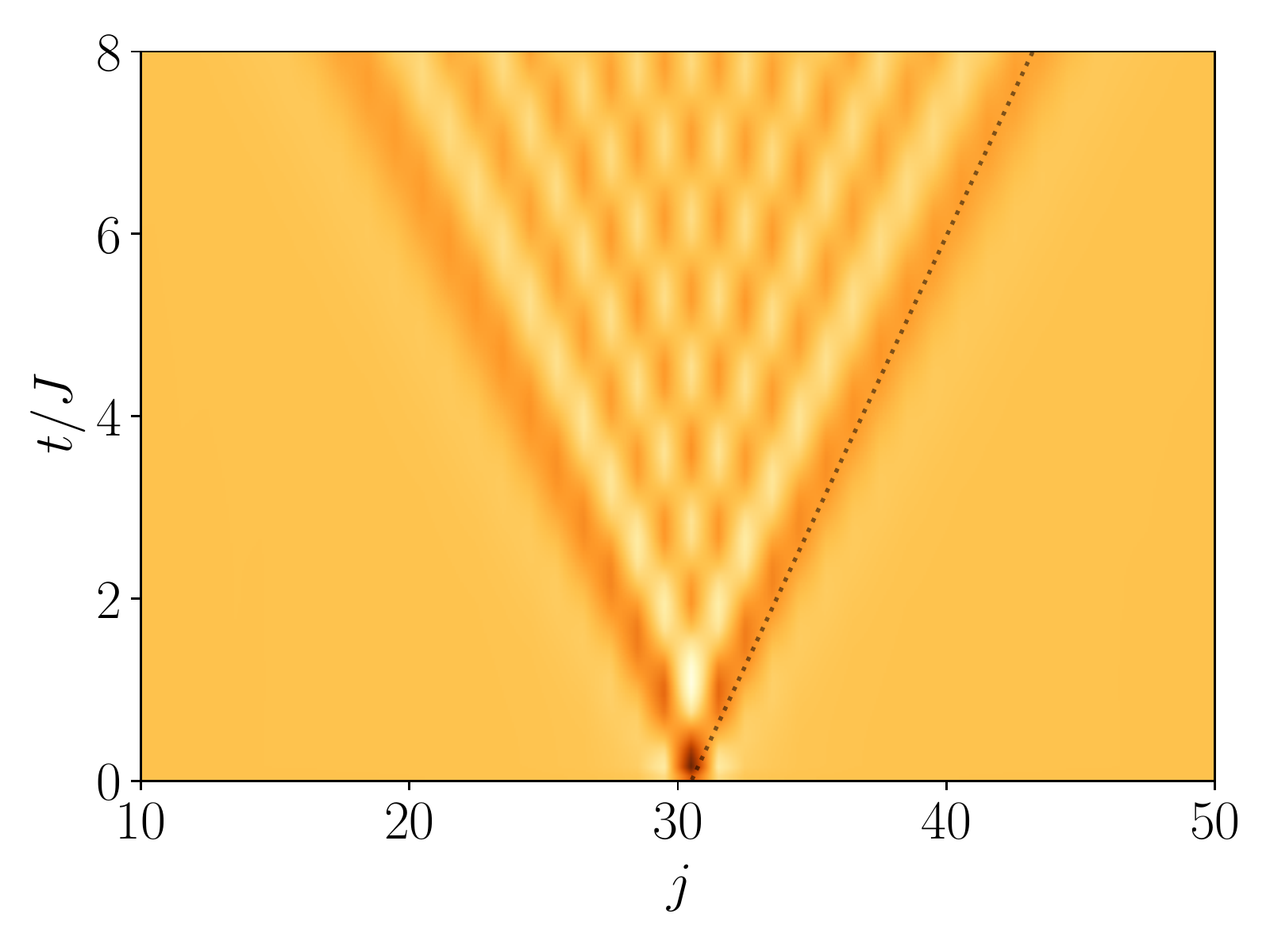}
\caption{A characteristic example of the space and time dependence of the correlation function $\langle \delta n_j(t) \delta n_0(t=0) \rangle$ at $U/J=20$ and $V/J=-0.5$ (GAS phase) with $N=30$ particles in a lattice of $N_s=60$ sites. The dotted line shows the slope given by the speed of sound extracted from the static structure factor according to Eq.~(\ref{Eq:Sk:Feynman}).}
\label{Fig:TimeEvol}
\end{figure}

To unravel the excitation spectrum of the system, we compute the dynamic structure factor $S(K,\omega)$. To this end, we calculate the space and time-dependent density-density correlation function in terms of $(x,t)$ variables and perform a Fourier transform in order to express it in conjugate $(k,\omega)$ variables
\begin{equation}
    S(k,\omega) = \int\limits_{-\infty}^{\infty}dt \sum_j \langle  \delta n_j(t) \delta n_0(0) \rangle\; e^{ikx_j-i\omega t},
\end{equation}
where $0$ index indicates the reference lattice site taken at the center of the system. To perform the time evolution, we employ the two-site TDVP algorithm~\cite{Haegeman2011,Haegeman2016}. We choose a time step $dt=0.1$ and a bond dimension $\chi = 800$. Since time evolution is not accurate at very long times, we employ the linear predictor to extrapolate our data to longer times~\cite{Pereira2008,Barthel2009}. Then, we use a Gaussian envelope. In Fig.~\ref{Fig:TimeEvol}, we show a typical time evolution obtained in the gas phase for the full model with dipolar interactions. The sound cone is formed, with an interference pattern within it. We find that the edge of the sound cone propagates with the same speed of sound $v_s$ as the one extracted through the lattice Feynman relation~(\ref{Eq:Sk:Feynman}).
\section{Lattice sum rules}
The dynamic structure factor obeys specific sum rules in the lattice~\cite{Roth2004}. The three main ones are given by,
\begin{eqnarray}
\label{Eq:sum rule:1}
m_1 &=& \int\, d\omega S(k,\omega) \omega = N_s \langle \hat{b}_i^{\dagger}\hat{b}_{i+1} \rangle \epsilon_k, \\
\label{Eq:sum rule:0}
m_0 &=& \int\, d\omega S(k,\omega) = N S(k), \\
\label{Eq:sum rule:-1}
m_{-1} &=& \int\, d\omega \frac{S(k,\omega)}{\omega} = \frac{N^2}{N_s} \frac{\kappa_s}{2}, 
\end{eqnarray}
where $\epsilon(k)=4J\sin (ka/2)^2$ is the single-particle dispersion relation in the lattice. Assuming  the energy spreading of the dynamic structure factor is small $\delta\omega = \omega_+-\omega_-\ll \omega$ and the dynamic structure factor is almost constant in that interval gives,
\begin{eqnarray}
S(k,\omega) \omega \delta\omega &=& N_s \langle \hat{b}_i^{\dagger}\hat{b}_{i+1} \rangle \epsilon(k), \\
S(k,\omega) \delta \omega &=& N S(k), \\
S(k,\omega)\frac{\delta \omega}{\omega} &=& \frac{N^2}{N_s} \frac{\kappa_s}{2}.
\end{eqnarray}
The above relations allow us to extract two possible Feynman relations by either using the zeroth and plus one ($f$-sum rule) 
or zeroth and minus one (compressibility) sum rules,
\begin{eqnarray}
\label{Eq:omega(k):Ekin}
\omega(k) &=& \frac{\langle \hat{b}_i^{\dagger}\hat{b}_{i+1} \rangle}{n} \frac{\epsilon(k)}{S(k)}, \\
\omega(k) &=& \frac{2}{n \kappa_s} S(k).
\label{Eq:omega(k):compressibility}
\end{eqnarray}
In the main text, we employed these Feynman relations. The first one can be applied to either compressible and incompressible phases. While the second one assumes a finite compressibility. Noteworthy, by using the compressibility~(\ref{Eq:sum rule:-1}) rule and assuming a linear phononic spectrum $\omega(k)=v_s k$ one recovers the Feynman relation~(\ref{Eq:Sk:Feynman}) which was derived using the non-Galilean invariant version of the Luttinger theory. 


Sum rules are also useful for the estimation of the weight of the dynamic structure factor. For a gapped phase $\omega(k) \sim \Delta$, we obtain,
\begin{align}
    S(k,\omega) \sim N_s \frac{\langle \hat{b}_i^{\dagger}\hat{b}_{i+1} \rangle}{ \delta w \Delta} \epsilon(k).
\end{align}
Moreover, in the gapped phase, the kinetic energy can be estimated $\langle \hat{b}_i^{\dagger}\hat{b}_{i+1} \rangle \sim n(n+1)/\Delta$ giving,
\begin{align}
    S(k,\omega) \sim  \frac{J N_s n(n+1)}{\delta w \Delta^2} \sin^2\left(\frac{ka}{2}\right).
\end{align}
Using the analytical expression for the lower and upper bounds of the excitation spectrum in the insulating phase~\cite{Iucci2006,Tokuno2011,Ejima2012,Ejima2012b},
\begin{align}
    w_\pm(k)=U\pm2J\sqrt{5+4\cos(ka)}, 
\end{align}
we obtain the dependence of the dynamic structure factor weight with the momenta,
\begin{align}
    S(k,\omega) \propto  \frac{\sin^2\left(\frac{ka}{2}\right)}{\sqrt{5+4\cos(ka)}}.
\end{align}
Thus, the weight vanishes quadratically at small momenta and reaches a maximum at the edge of the Brillouin zone. This result is compatible with our numerical results in the B-MI phase. 

At small dipolar interaction the bounds of the excitation spectrum can be estimated from the Bethe ansatz solution at zero dipolar strength~\cite{Cloizeaux1962,Yamada1969},
\begin{align}
    w_+(k)=4J \big|\sin\left(\frac{ka}{2} \right)\big|, \quad  w_-(k)=2J |\sin(ka)|.
\end{align}
This gives a dynamic structure factor that diverges at small momenta,
\begin{align}
    S(k,w)&=N_s \langle \hat{b}_i^{\dagger}\hat{b}_{i+1} \rangle \frac{\epsilon(k)}{\delta \omega(k) \omega(k)}\\
    &=\frac{N_s \langle \hat{b}_i^{\dagger}\hat{b}_{i+1}\rangle}{ v_s}  \left(\frac{2}{ka}\right)^2 + \mathcal{O}(1).
\end{align}

\section{Experimental implementation}
Ultracold atoms provide an ideal experimental platform to realize a strongly interacting lattice bosonic system with an attractive long-range interaction. Specifically, dipolar bosons loaded to a one-dimensional optical lattice are perfect candidates to produce a long-range interaction with a power-law decay.
One can obtain a pure dipolar interaction by assuming that the characteristic transverse length $\sigma_{\perp}$ is much smaller than the lattice spacing in the longitudinal direction $a_x$, which is fulfilled in recent experiments~\cite{Kao296}. Moreover, the on-site interaction can also be tuned by employing Feshbach resonances. Finally, the strength of the dipolar interaction can be tuned by changing the polarization angle $\theta$ between the dipoles,
\begin{align}
    V = \frac{C_{dd}}{4\pi} \frac{1-3\cos^2(\theta)}{a_x^3},
\end{align}
being $C_{dd}$ the dipolar coupling.

Considering the recent quasi one-dimensional setting for $^{162}$Dy atoms~\cite{Kao296} ($C_{dd}\approx (9.93\mu_B)^2 \mu_0$) with $\sigma_{\perp}= 952 a_0$, $a_0$ being the Bohr radius, $V_{\perp}=30 E_{\perp}$ and $E_{\perp}/\hbar=2\pi \times 2.24$kHz and a longitudinal optical lattice of lattice spacing $a_x=532$nm with a height $V_x=14 E_x$, one can cross all the phases encountered in the main text by changing the polarization angle.

\end{document}